\documentclass[aps,prl,superscriptaddress,twocolumn,longbibliography]{revtex4-1}

\usepackage{amsfonts}
\usepackage{subfigure}
\usepackage{amsmath}
\usepackage{txfonts}
\usepackage{bm}
\usepackage{mathrsfs}
\usepackage{amssymb}
\usepackage{amsbsy} 
\usepackage{epsfig}
\usepackage{graphicx}
\usepackage{color}
\usepackage{epstopdf}

\def\be{\begin{equation}} \def\ee{\end{equation}}
\def\bea{\begin{eqnarray}} \def\eea{\end{eqnarray}}

\def\nn{\nonumber}

\def\bk{{\bf k}}

\def\bp{{\bf p}}

\def\be{{\bf e}}

\def\ra{\rangle}

\def\rw{\rightarrow}

\begin{document}

\title{Majorana Corner Modes in a High-Temperature Platform}

\author{Zhongbo Yan}
 \affiliation{ Institute for
Advanced Study, Tsinghua University, Beijing,  100084, China }

\author{Fei Song}
 \affiliation{ Institute for
Advanced Study, Tsinghua University, Beijing,  100084, China }

\author{Zhong Wang} \altaffiliation{ wangzhongemail@gmail.com }
\affiliation{ Institute for
Advanced Study, Tsinghua University, Beijing,  100084, China }

\affiliation{Collaborative Innovation Center of Quantum Matter, Beijing, 100871, China }

%

\begin{abstract}

We introduce two-dimensional topological insulators in proximity to high-temperature cuprate or iron-based superconductors as high-temperature platforms of Majorana Kramers pairs of zero modes. The proximity-induced pairing at the helical edge states of topological insulator serves as a Dirac mass, whose sign changes at the sample corner because of the pairing symmetry of high-$T_c$ superconductors. This sign changing naturally creates at each corner a pair of Majorana zero modes protected by time-reversal symmetry. Conceptually, this is a topologically-trivial-superconductor-based approach for Majorana zero modes. We provide quantitative criteria and suggest candidate materials for this proposal.

\end{abstract}


\maketitle

Majorana zero modes (MZMs)\cite{read2000,volovik1999fermion,kitaev2001unpaired} have been actively pursued in recent years as building blocks of topological quantum computations\cite{alicea2012new,Beenakker2013,stanescu2013majorana, leijnse2012introduction,Elliott2015,sarma2015majorana,sato2016majorana,aguado2017}. These emergent excitations can generate robust ground-state degeneracy, supporting storage of nonlocal qubits robust to local decoherence\cite{nayak2008}. Moreover, quantum gates can be implemented by their braiding operations\cite{moore1991nonabelions,wen1991a,ivanov2001,nayak1996,sarma2005}.
As platforms of MZMs, a variety of realizations of topological superconductors have been proposed, including
topological insulators in proximity to conventional superconductors\cite{fu2007c,qi2010chiral,akhmerov2009,law2009majorana,Chung2011}, semiconductor heterostructures\cite{sau2010,alicea2010,Xu2014TSC}, cold-atom systems\cite{Jiang2011,diehl2011topology,sato2009non,zhang2008px,tewari2007quantum,liu2014realization}, quantum wires\cite{oreg2010helical,lutchyn2010,vazifeh2013,cook2011}, to name a few; meanwhile, remarkable experimental progress has been witnessed\cite{mourik2012signatures, nadj2014observation,rokhinson2012fractional,deng2012anomalous, das2012zero,finck2013,churchill2013superconductor,lv2016,Wang2012coexistence,wang2016topological, albrecht2016exponential,Deng2016Majorana,pawlak2015probing, Xu2015experimental, Sun2016Majorana,he2016chiral,zhang2017quantized,zhang2018iron,wang2017iron}.

A single MZM entails breaking the time-reversal symmetry (TRS); in contrast, time-reversal-invariant (TRI) topological superconductors\cite{qi2009b,qi2010d,zhang2013kramers,Wong2012majorana,sato2010odd, Zhang2013mirror,fu2010odd,haim2016} host Majorana Kramers pairs (MKPs) of zero modes, which are robust in the presence of TRS, and have interesting consequences such as TRS-protected non-Abelian statistics\cite{liu2014non,Zhang2014anomalous,gao2016symmetry}, TRS as local supersymmetry\cite{qi2009b}, and novel Kondo effects\cite{Bao2017kondo} and Josephson effects \cite{zhang2014,Orth2015,Schrade2015,schrade2018parity,Camjayi2017fractional}, indicating their potentials in qubit storage or manipulation and other applications. In addition, MKPs can be used as tunable generators of MZMs by breaking the TRS\cite{zhang2013kramers,Wong2012majorana}. There have been a few interesting proposals for realizing TRI topological superconductors and MKPs \cite{zhang2013kramers,Wong2012majorana,keselman2013,nakosai2013,Nakosai2012, wang2014TRI,haim2014,Klinovaja2014,yang2015time,Trani2016, li2016detection,wu2017majorana,Reeg2017,wang2016electrically,hu2017doublets}, though experimental realizations are yet to come.

\begin{figure}
\subfigure{\includegraphics[width=7cm, height=2.6cm]{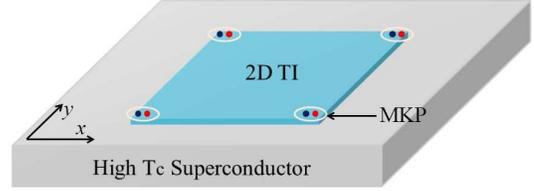}}
\caption{Schematic illustration. A 2D TI is grown on
a $d$-wave or $s_\pm$-wave high-$T_{c}$ superconductor. Majorana Kramers pairs (MKPs) of zero modes emerge at the corners of TI. }  \label{sketch}
\end{figure}

In this Letter, we show that simple structures of two-dimensional topological insulators (2D TIs) (also known as quantum spin Hall insulators) in proximity to high-temperature superconductors naturally generate MKPs (Fig.\ref{sketch}). Since 2D TIs have been experimentally realized at temperatures as high as 100 Kelvin\cite{Wu2018QSHE,Qian2014TMD}, this setup can be a high-temperature platform of MKPs. The physical picture can be readily described as follows. The helical edge states of TI, described as 1D massless Dirac fermions, are gapped out by the induced superconducting gap, which introduces a Dirac mass. Due to the nature of pairing symmetry (say $d$-wave), the induced Dirac mass changes sign at the corner, which generates a MKP as domain-wall excitations.

It is interesting to note that we do not propose here any realization of TRI topological superconductor. In fact, the helical Majorana edge states of $\mathbb{Z}_2$-nontrivial superconductors cannot be gapped out without breaking TRS. In our setup, the 2D TI with a proximity-induced pairing has gapped edges, therefore it is a $\mathbb{Z}_2$-trivial superconductor. In fact, it has recently been suggested that, as defect modes\cite{teo2010}, robust MZMs can be realized in certain topologically trivial superconductors\cite{MZM-Hopf,chan2017generic} (Particularly, MZMs can in principle be created as corner modes in judiciously designed trivial-superconductor junctions\cite{MZM-Hopf}). Conceptually, the present work generalizes the trivial-superconductor-based approach to MKPs, for which ideal candidate materials are available.

\emph{$d$-wave pairing.--}As explained above, the key observation comes from the edge states. For concreteness, however, let us start from a lattice model of 2D TI, in which the proximity-induced pairing is added. The Bogoliubov-de Gennes Hamiltonian is $\hat{H}=\sum_{\bk}\Psi_{\bk}^{\dag}H(\bk)\Psi_{\bk}$,
with $\Psi_{\bk}=(c_{a,\bk\uparrow},c_{b,\bk\uparrow}, c_{a,\bk\downarrow},c_{b,\bk\downarrow},
c_{a,-\bk\uparrow}^{\dag},c_{b,-\bk\uparrow}^{\dag}, c_{a,-\bk\downarrow}^{\dag},c_{b,-\bk\downarrow}^{\dag})^{T}$ and
\begin{eqnarray}
H(\bk)=&&M(\bk)\sigma_{z}\tau_{z}+A_{x}\sin k_{x}\sigma_{x}s_{z}+A_{y}\sin k_{y}\sigma_{y}\tau_{z} \nn \\ && +\Delta(\bk)s_{y}\tau_{y}-\mu\tau_z,\quad\label{lattice}
\end{eqnarray}
where $s_{i}$, $\sigma_{i}$ and $\tau_{i}$ are Pauli matrices in the
spin ($\uparrow,\downarrow$), orbital ($a,b$), and particle-hole space, respectively,
$M(\bk)= m_{0}-t_{x}\cos k_{x}-t_{y}\cos k_{y}$ and $A_{x,y}$ measure the kinetic energy, $\Delta$ is the pairing, and $\mu$ is the chemical potential. In the following we will take \bea \Delta(\bk)=\Delta_{0}+\Delta_{x}\cos k_{x}+\Delta_{y}\cos k_{y},\eea which is sufficiently general to model $d$ wave and $s_\pm$ wave.
Throughout this paper, $t_{x,y}$, $A_{x,y}$ are taken to be positive. If the pairing is removed, the Hamiltonian becomes the paradigmatic BHZ model of 2D TIs\cite{bernevig2006c,Qian2014TMD,Wu2018QSHE}.
The Hamiltonian has TRS $\mathcal{T}H(\bk)\mathcal{T}^{-1}=H(-\bk)$ with
$\mathcal{T}=is_{y}\mathcal{K}$ (where $\mathcal{K}$ is the complex conjugation), and
particle-hole symmetry $\mathcal{C}H(\bk)\mathcal{C}^{-1}=-H(-\bk)$ with
$\mathcal{C}=\tau_{x}\mathcal{K}$.

\begin{figure}
\subfigure{\includegraphics[width=4.2cm, height=4.2cm]{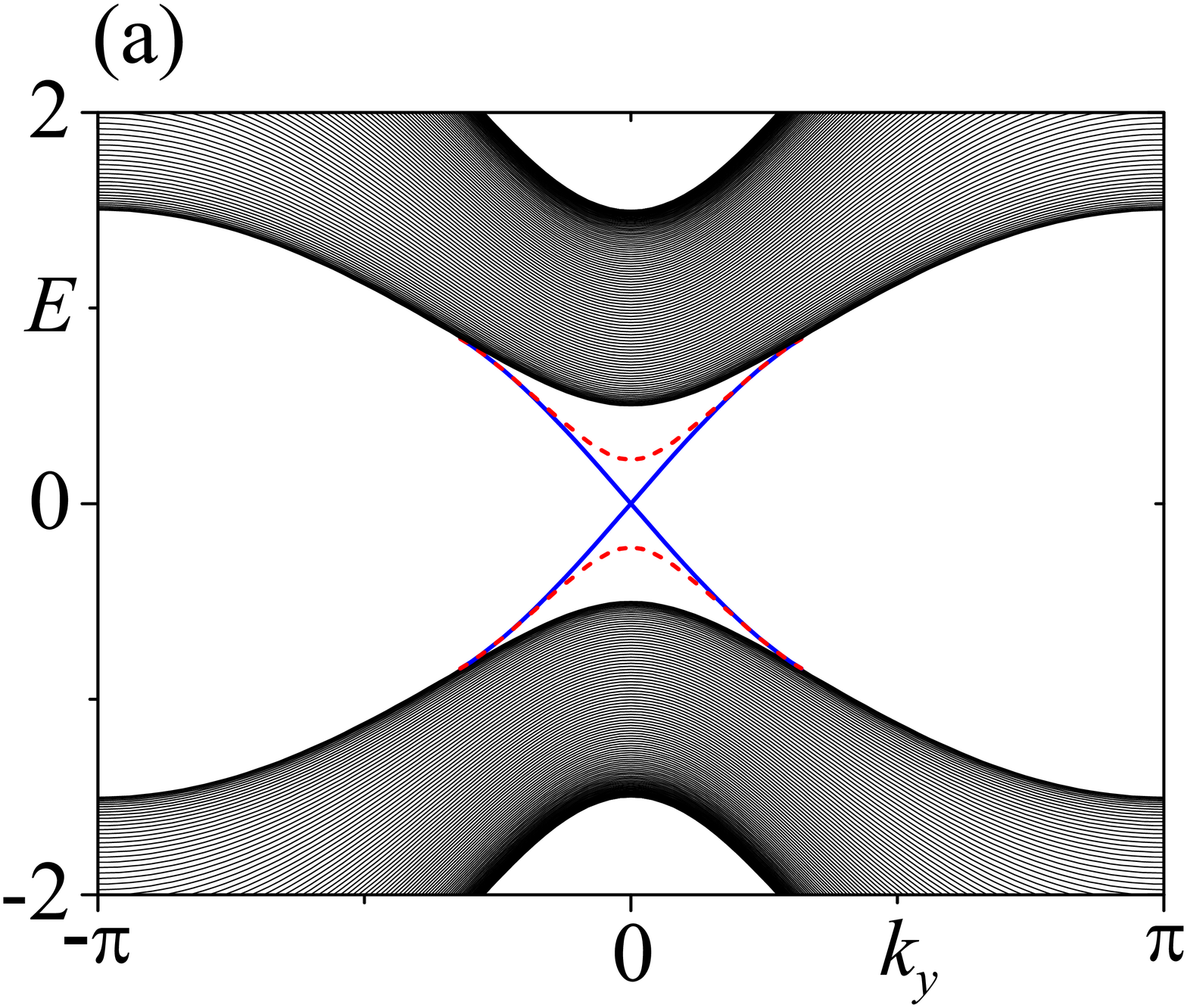}}
\subfigure{\includegraphics[width=4.2cm, height=4.2cm]{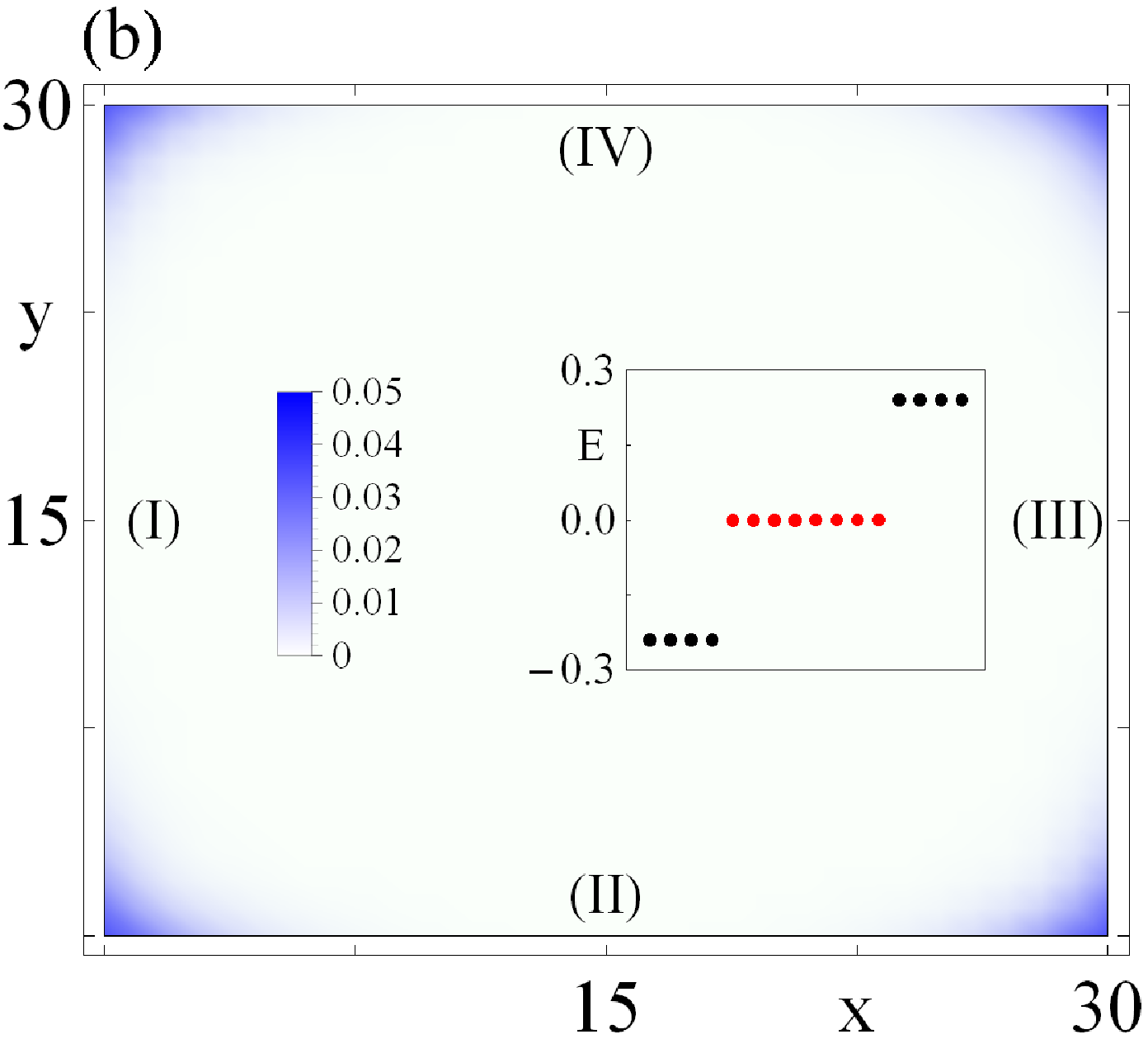}}
\caption{(a) Energy spectra in a cylinder geometry. $m_{0}=1.5$, $t_{x}=t_{y}=1.0$, $A_{x}=A_{y}=1.0$, $\mu=0$.
Without the pairing, there exist helical edge states traversing the bulk gap (solid blue lines).
In the presence of a $d$-wave pairing ($\Delta_x=-\Delta_y=0.5$), the edge states become gapped (dashed red lines). The bulk spectra have little difference for these two cases (the zero-pairing case is shown here).
(b) The wavefunction profiles of the four MKPs from solving the real-space lattice Hamiltonian. The sample size is $L_{x}\times L_{y}=30\times30$. The inset shows energies near zero, indicating one MKP per corner. (I), (II), (III), and (IV) mark the four edges for use in the edge theory.  }  \label{MZM}
\end{figure}

We first consider the $d$-wave pairing that is relevant to cuprate superconductors, which is \bea \Delta_{0}=0, \quad \Delta_{x}=-\Delta_{y}\equiv  \Delta_d. \eea The spectra on a cylinder geometry are shown in Fig.\ref{MZM}(a), indicating that the helical edge states of TI are gapped out by $d$-wave pairing. From the numerical results for a square geometry [Fig.\ref{MZM}(b)], it is clear that each corner hosts a MKP, whose energy is pinned to zero.

It is interesting to note that, unlike the more familiar vortex or end modes, the Majorana modes here are corner modes. As such, they may be viewed in the framework of recently proposed higher-order topological insulators\cite{benalcazar2017quantized,Schindler2017higher,Zhang2013surface, Benalcazar2017prb,Song2017higher,Langbehn2017hosc,peng2017boundary,
Imhof2017corner,serra2018,Schindler2018bismuth,ezawa2018higher,Peterson2018} and superconductors\cite{Geier2018hosc,Zhu2018hosc,Khalaf2018hosc,Shapourian2018}, for which crystal symmetries have been highlighted; the present scheme does not rely sensitively on the crystal symmetries.

We also mention that the bulk of $d$-wave superconductor is gapless and the MKPs may hybridize with these gapless modes. Nevertheless, the MKPs remain observable in scanning
tunneling microscopy (STM). Near a MKP, the tunneling conductance displays a zero-bias peak (though broadened by hybridization), which is absent in the usual $c$-axis tunneling conductance\cite{Fischer2007} (In other directions there is zero bias peak \cite{Kashiwaya1995,Sinha1998,Tanaka1995,Ryu2002,Shigeta2002,Giubileo2002,Kao2015}, which is irrelevant in our setup).  Later, we also study the $s_\pm$-wave case with an entirely gapped setup, in which case MKPs are the only low-energy modes.

{\it Edge theory.--}To gain intuitive understandings, we study the edge theory. To simplify the picture, we take $\mu=0$ and focus on the continuum model by expanding the lattice Hamiltonian in Eq.(\ref{lattice}) to second order around $\bk=(0,0)$:
\begin{eqnarray}
H(\bk)&=&(m+\frac{t_{x}}{2}k_{x}^{2}+\frac{t_{y}}{2}k_{y}^{2})\sigma_{z}\tau_{z}+A_{x}k_{x}\sigma_{x}s_{z}+A_{y}k_{y}\sigma_{y}\tau_{z}\nonumber\\
&&-\frac{1}{2}(\Delta_{x} k_{x}^{2}+\Delta_y k_{y}^{2})s_{y}\tau_{y},\label{continuum}
\end{eqnarray}
where $\Delta_x+\Delta_y=0$ has been used for the $d$ wave, and $m=m_{0}-t_{x}-t_{y}<0$ is assumed to ensure that the 2D insulator without pairing is in the topologically nontrivial regime. We label the four edges of a square as (I), (II), (III), and (IV) [Fig.\ref{MZM}(b)], and we focus on the edge (I) first. We can replace $k_x\rw -i\partial_x$ and decompose the Hamiltonian as $H=H_{0}+H_{p}$, in which
\begin{eqnarray}
H_{0}(-i\partial_{x},k_{y})&=&(m-t_x\partial_{x}^{2}/2)\sigma_{z}\tau_{z}-iA_{x}\sigma_{x}s_{z}\partial_{x},\nonumber\\
H_{p}(-i\partial_{x},k_{y})&=&A_{y} k_{y}\sigma_{y}\tau_{z}+(\Delta_{x}/2)s_{y}\tau_{y}\partial_{x}^{2},
\end{eqnarray} where the insignificant $k_y^2$ term has been omitted. The purpose of this decomposition is to solve $H_0$ first, and then treat $H_p$ as a perturbation, which is justified when the pairing is relatively small (This is the case in real samples).

Solving the eigenvalue equation $H_{0}\psi_{\alpha}(x)=E_{\alpha}\psi_{\alpha}(x)$ under the boundary condition
$\psi_{\alpha}(0)=\psi_{\alpha}(+\infty)=0$, we find four zero-energy solutions, whose forms are
\bea \psi_{\alpha}(x)=\mathcal{N}_{x}\sin(\kappa_{1}x) e^{-\kappa_{2}x}e^{ik_{y}y}\chi_{\alpha},\eea
with normalization given by $|\mathcal{N}_{x}|^2=4|\kappa_{2}(\kappa_{1}^{2}+\kappa_{2}^{2})/\kappa_{1}^{2}|$ (Here, $\kappa_{1}=\sqrt{ |\frac{2m}{t_{x}}|-\frac{A_{x}^{2}}{t_{x}^{2}}}, \,\kappa_{2}=\frac{A_{x}}{t_{x}}$. The result remains valid even when $\kappa_1$ is imaginary). The eigenvectors $\chi_{\alpha}$  satisfy $\sigma_{y}s_{z}\tau_{z}\chi_{\alpha}=-\chi_{\alpha}$.
We can explicitly choose them as \bea \chi_{1}=|\sigma_{y}=-1\rangle\otimes|\uparrow\rangle\otimes|\tau_{z}=+1\rangle,\nn\\
\chi_{2}=|\sigma_{y}=+1\rangle\otimes|\downarrow\rangle\otimes|\tau_{z}=+1\rangle,\nn\\
\chi_{3}=|\sigma_{y}=+1\rangle\otimes|\uparrow\rangle\otimes|\tau_{z}=-1\rangle,\nn\\
\chi_{4}=|\sigma_{y}=-1\rangle\otimes|\downarrow\rangle\otimes|\tau_{z}=-1\rangle, \eea then the matrix elements of the perturbation $H_p$ in this basis are
\begin{eqnarray}
H_{{\rm I}, \alpha\beta}(k_{y})=\int_{0}^{+\infty} dx\psi^*_{\alpha}(x)H_{p}(-i\partial_{x},k_{y})\psi_{\beta}(x),
\end{eqnarray}
therefore, the final form of the effective Hamiltonian is
\begin{eqnarray}
H_{\rm I}(k_{y})=-A_{y} k_{y}s_{z}+M_{\rm I}s_{y}\tau_{y},
\end{eqnarray}
where \bea M_{\rm I}=(\Delta_{x}/2)\int_{0}^{+\infty} dx\psi_{\alpha}^*(x) \partial_{x}^{2}\psi_{\alpha}(x)= \Delta_{x}m/t_{x}. \eea
Similarly, the low-energy effective Hamiltonians for the other three edges are
\begin{eqnarray}
H_{\rm II}(k_{x})&=&A_{x} k_{x}s_{z}+M_{\rm II}s_{y}\tau_{y},\nonumber\\
H_{\rm III}(k_{y})&=&A_{y} k_{y}s_{z}+M_{\rm III}s_{y}\tau_{y},\nonumber\\
H_{\rm IV}(k_{x})&=&-A_{x} k_{x}s_{z}+M_{\rm IV}s_{y}\tau_{y} \label{otherthree}
\end{eqnarray}
with $M_{\rm II}=M_{\rm IV}= \Delta_ym/t_{y}$, and $M_{\rm III}=M_{\rm I}$. To be more transparent, let us take an ``edge coordinate'' $l$, which grows in the anticlockwise direction (apparently, $l$ is defined mod $2(L_x+L_y)$), then the low-energy edge theory becomes
\begin{eqnarray}
H_{\rm edge}=-iA(l)s_{z}\partial_{l}+M(l)s_{y}\tau_{y}. \label{edge-H}
\end{eqnarray}
The kinetic-energy coefficient $A(l)$ and the Dirac mass $M(l)$ are step functions: $A(l)=A_y,A_x,A_y,A_x$ and $M(l)= \Delta_d m/t_x,- \Delta_d m/t_y, \Delta_d m/t_x,- \Delta_d m/t_y$ for (I), (II), (III), and (IV), respectively. At each corner, the $A_{x,y}$ coefficient does not change sign, while the Dirac mass does, which is due to the sign changing in the $d$-wave pairing: $\Delta_x=-\Delta_y$. Consequently, there is a MKP at each corner (analogous to the Jackiw-Rebbi zero modes\cite{jackiw1976b,lee2007}). For example, at the corner between (I) and (II), we have \bea |\psi_\text{MKP}^\pm(l)\ra \propto e^{ -\int^l dl' M(l')/A(l')} | s_x=\tau_y=\pm1 \ra. \eea   TRS ensures that these two modes cannot be coupled to generate an energy gap.
In essence, the edge theory above can be regarded as two copies of that of Ref.\cite{MZM-Hopf}, with TRS as the key additional input.

\begin{figure}
\subfigure{\includegraphics[width=7cm, height=5cm]{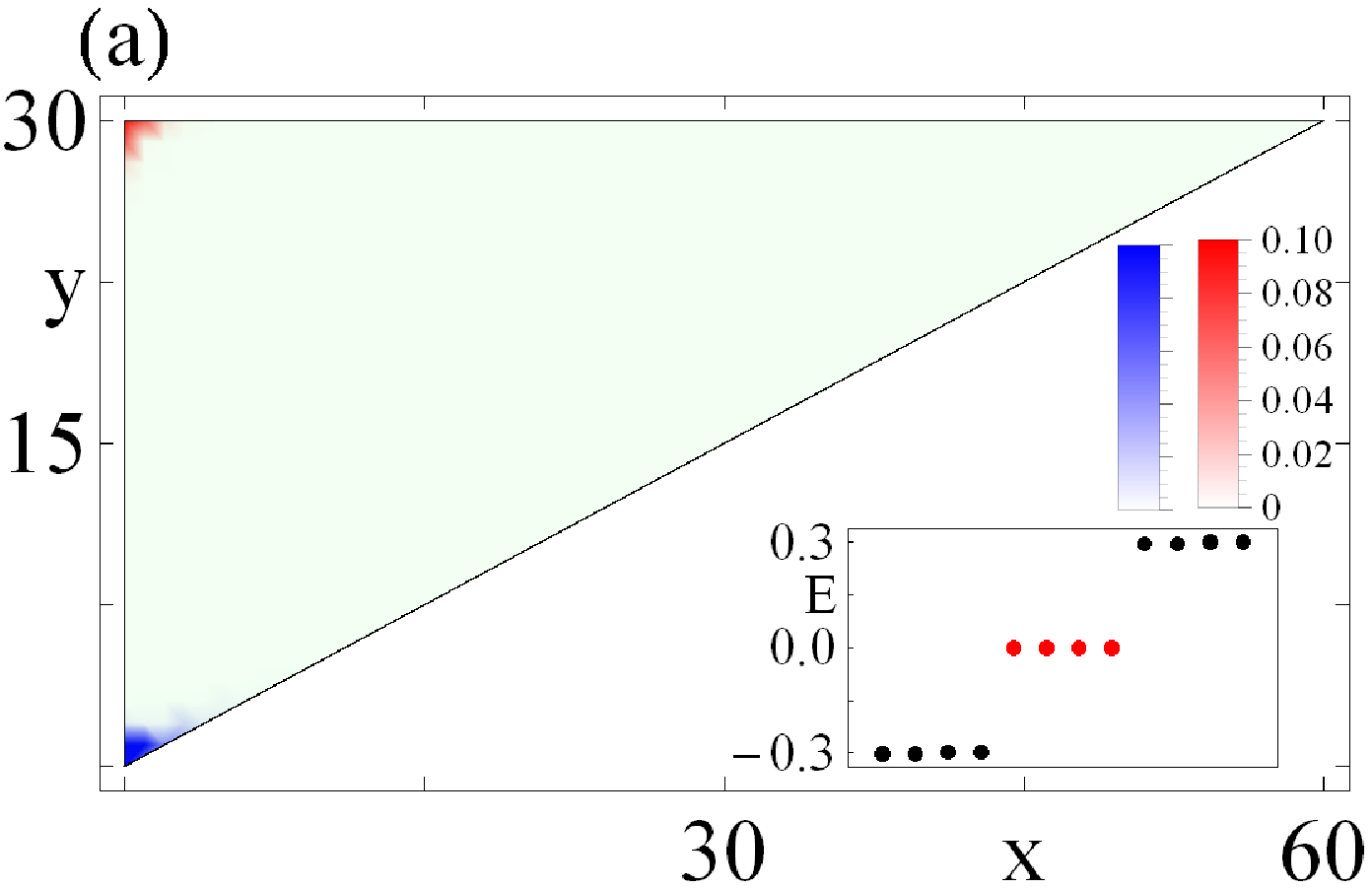}}
\subfigure{\includegraphics[width=5.5cm, height=5cm]{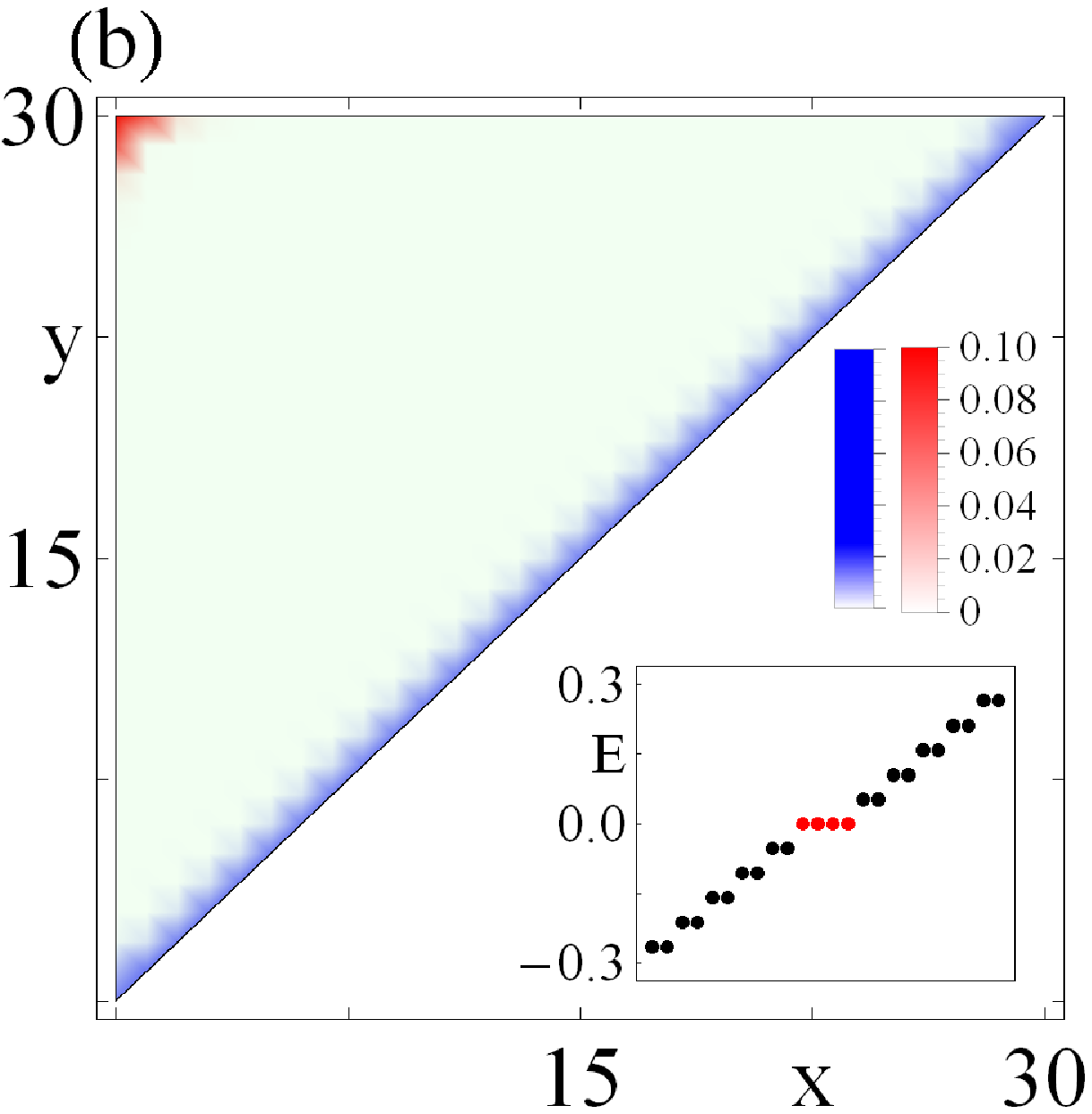}}
\caption{ MKPs in triangle samples. (a) The existence or absece of MKPs depends on the edge directions at the corner, which can be explained in the edge theory. The lower corner has sign change in the edge Dirac mass, while the right corner does not. (b) For a $\pi/4$ angle, the edge Dirac mass vanishes, and the edge states display a gapless feature (see the inset, and compare it to that of (a)).    $m_{0}=1.5$, $t_{x}=t_{y}=2.0$, $A_{x}=A_{y}=2.0$,
$\Delta_{x}=-\Delta_y=1.0$, $\mu=0$. } \label{triangle}
\end{figure}

By a similar calculation, one can find that the sign changing in $M(l)$ occurs at a corner when one of the edges has a polar angle within $[-\pi/4,\pi/4]$ and the other within $[\pi/4,3\pi/4]$ (the gap-maximum direction is taken as the zero polar angle). In Fig.\ref{triangle}(a), the lower corner has a sign changing while the right corner does not, and the existence or absence of MKP is consistent with the edge-theory prediction. If one of the edges lies in the $\pi/4$ direction, the edge states become gapless, which also manifests in the numerical spectrum in Fig.\ref{triangle}(b).

Finally, we mention that cuprate superconductors in proximity to 3D topological insulators have been experimentally studied for the purpose of creating vortex (instead of corner) MZMs\cite{Wang2013proximity,Zareapour2012proximity,li2015realizing,Ortiz2018majorana}. In these setups, the 2D topological surface states (instead of the 1D edge states) are the key ingredients.

\emph{$s_\pm$-wave pairing.--}Now we consider fully gapped $s_\pm$-wave superconductors with sign changing in the pairing.
A host of candidates can be found in high $T_c$ iron-based superconductor\cite{stewart2011superconductivity,hirschfeld2011gap}, whose pairing at the Fermi surfaces near the Brillouin zone
center and the Brillouin zone boundary have both $s$-wave nature but with opposite signs. The Fermi surfaces do not cross the pairing nodal rings, therefore, the superconductor is fully gapped.
A simplest form of $s_\pm$-wave pairing is: \bea \Delta(\bk)=\Delta_{0}-\Delta_{1}(\cos k_{x}+\cos k_{y}),\eea with $0<\Delta_{0}<2\Delta_{1}$. The pairing node is $\cos k_x+\cos k_y= \Delta_0/\Delta_1$.

Let us first study the edge theory of TI.
Expanding the Hamiltonian near $\bk=(0,0)$ and taking $\mu=0$, we have
\begin{eqnarray}
H(\bk)&=&(m+\frac{t_{x}}{2}k_{x}^{2}+\frac{t_{y}}{2}k_{y}^{2})\sigma_{z}\tau_{z}+A_{x}k_{x}\sigma_{x}s_{z}+A_{y}k_{y}\sigma_{y}\tau_{z}\nonumber\\
&&+[\Delta_{0}-2\Delta_{1}+\frac{\Delta_{1}}{2}(k_{x}^{2}+k_{y}^{2})]s_{y}\tau_{y}.
\end{eqnarray}
Following a similar approach as the previous section, for the edge (I), we decompose the Hamiltonian as $H=H_{0}+H_{p}$, where
\begin{eqnarray}
H_{0}(-i\partial_{x},k_{y})&=&(m-t_{x}\partial_{x}^{2}/2)\sigma_{z}\tau_{z}-iA_{x}\sigma_{x}s_{z}\partial_{x},\nonumber\\
H_{p}(-i\partial_{x},k_{y})&=&A_{y} k_{y}\sigma_{y}\tau_{z}+[\Delta_{0}-2\Delta_{1}-(\Delta_{1}/2)\partial_{x}^{2}]s_{y}\tau_{y}.
\end{eqnarray} Similar to the previous section, four zero-energy solutions of $H_0$ can be found, and $H_p$ takes the following form within this four-dimensional low-energy subspace:
\begin{eqnarray}
H_\text{I}(k_{y})&=&-A_{y} k_{y}s_{z}+M_\text{I}s_{y}\tau_{y},
\end{eqnarray}
with $M_\text{I}=\int_{0}^{+\infty} dx\psi^{*}_{\alpha}(x)[\Delta_{0}-2\Delta_{1}-(\Delta_{1}/2)\partial_{x}^{2}]\psi_{\alpha}(x) =\Delta_{0}-2\Delta_{1}-\Delta_{1}m/t_{x}$. The low-energy effective Hamiltonians for the other three edges take the same forms as in Eq.(\ref{otherthree}), with Dirac masses $M_\text{III}=\Delta_{0}-2\Delta_{1}-\Delta_{1}m/t_{x}=M_\text{I}$, and
$M_\text{II}=M_\text{IV}=\int_{0}^{+\infty} dy\psi^*_{\alpha}(y) [\Delta_{0}-2\Delta_{1}-(\Delta_{1}/2)\partial_{y}^{2}]\psi_{\alpha}(y)=\Delta_{0}-2\Delta_{1}-\Delta_{1}m/t_{y}$. Using the edge coordinate $l$, the effective edge Hamiltonian is the same as Eq.(\ref{edge-H}) with the same $A(l)$ but different $M(l)$, namely, $M(l)=-\bar{\Delta}_0-\Delta_1 m/t_x, -\bar{\Delta}_0-\Delta_1 m/t_y, -\bar{\Delta}_0-\Delta_1 m/t_x, -\bar{\Delta}_0-\Delta_1 m/t_y$ for (I), (II), (III), and (IV), respectively, where we have defined $\bar{\Delta}_0=2\Delta_1-\Delta_0$.

\begin{figure*}[t!]
\subfigure{\includegraphics[width=4.0cm, height=4.0cm]{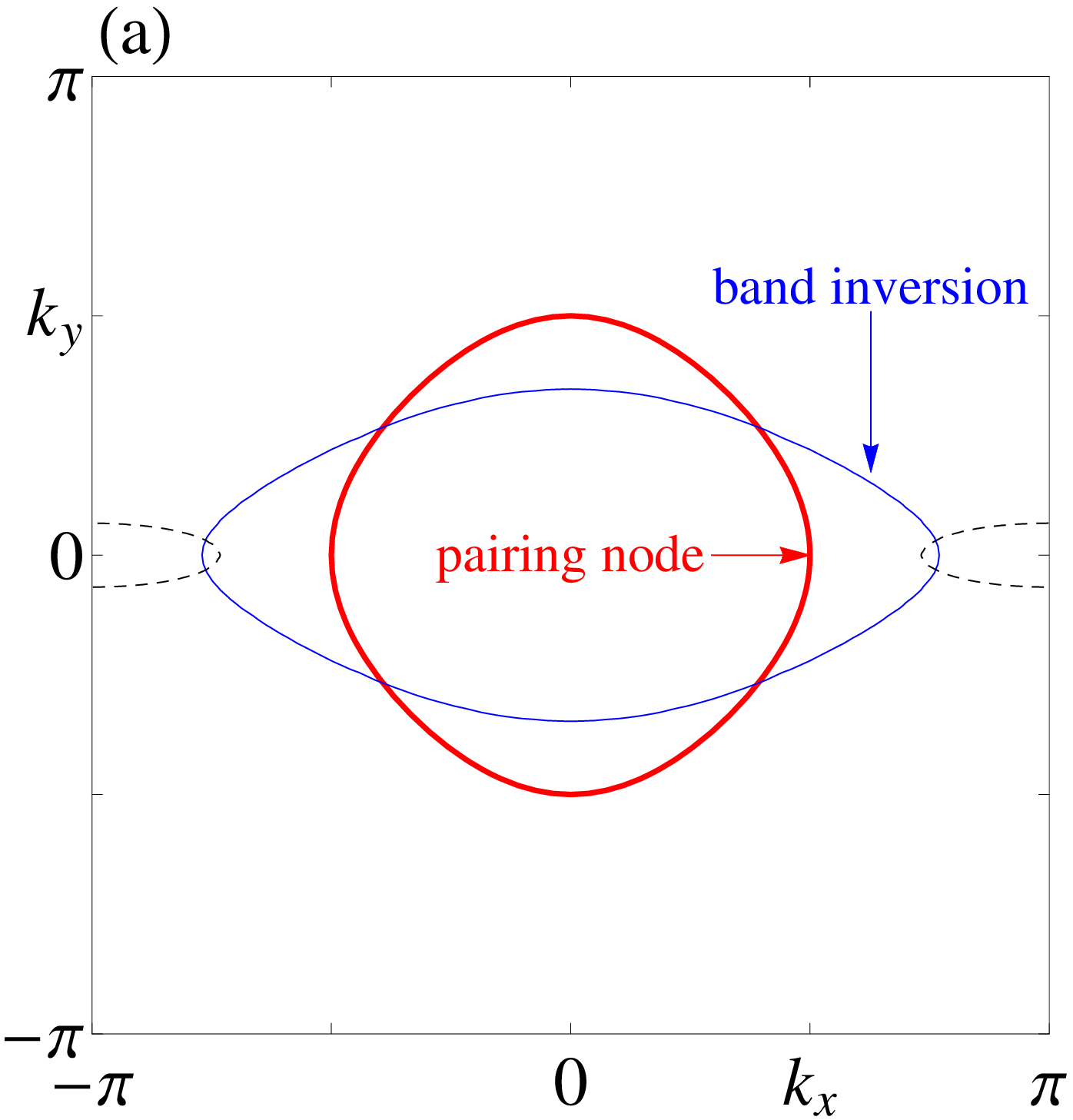}}
\subfigure{\includegraphics[width=4.0cm, height=4.0cm]{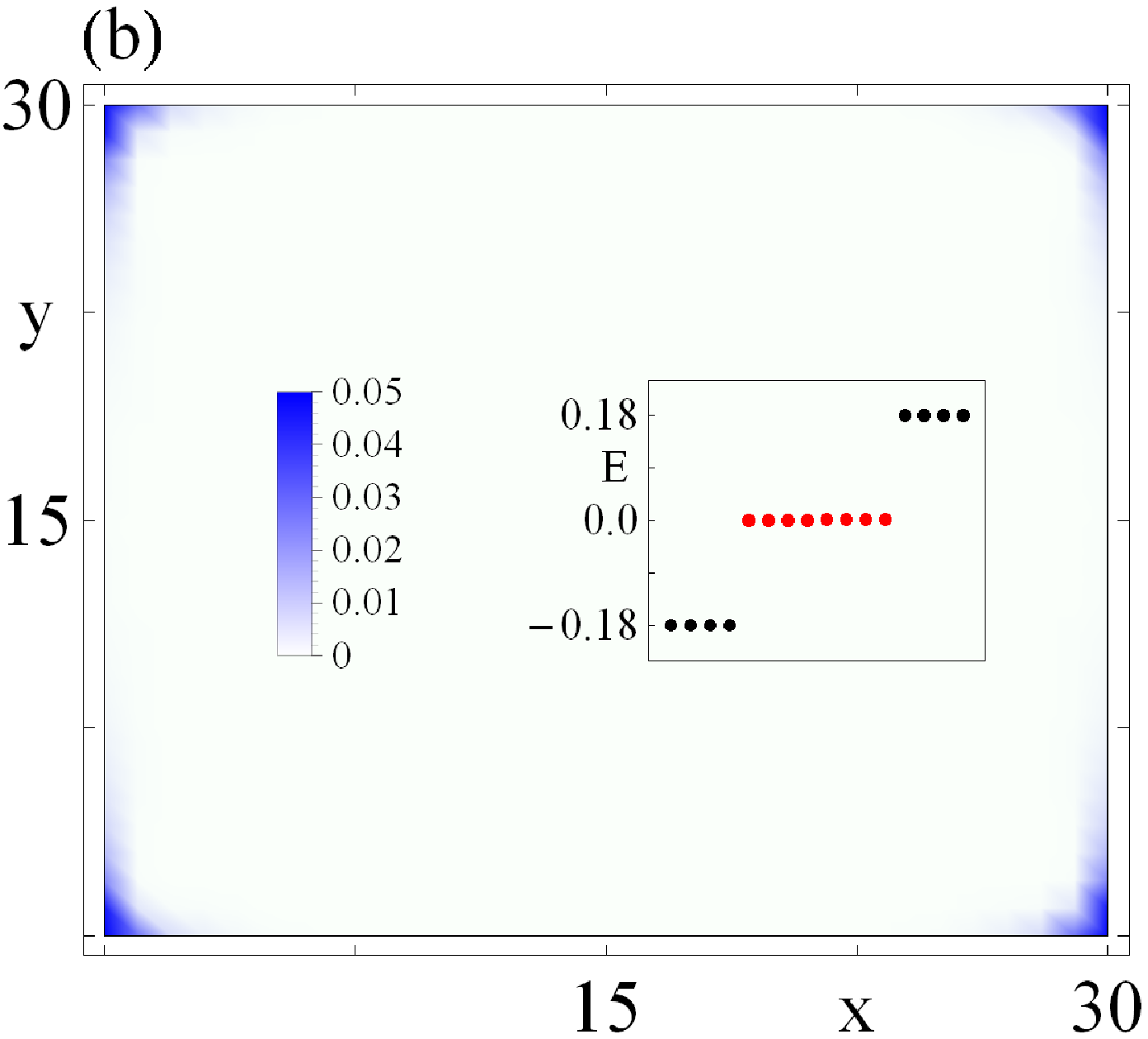}}
\subfigure{\includegraphics[width=4.0cm, height=4.0cm]{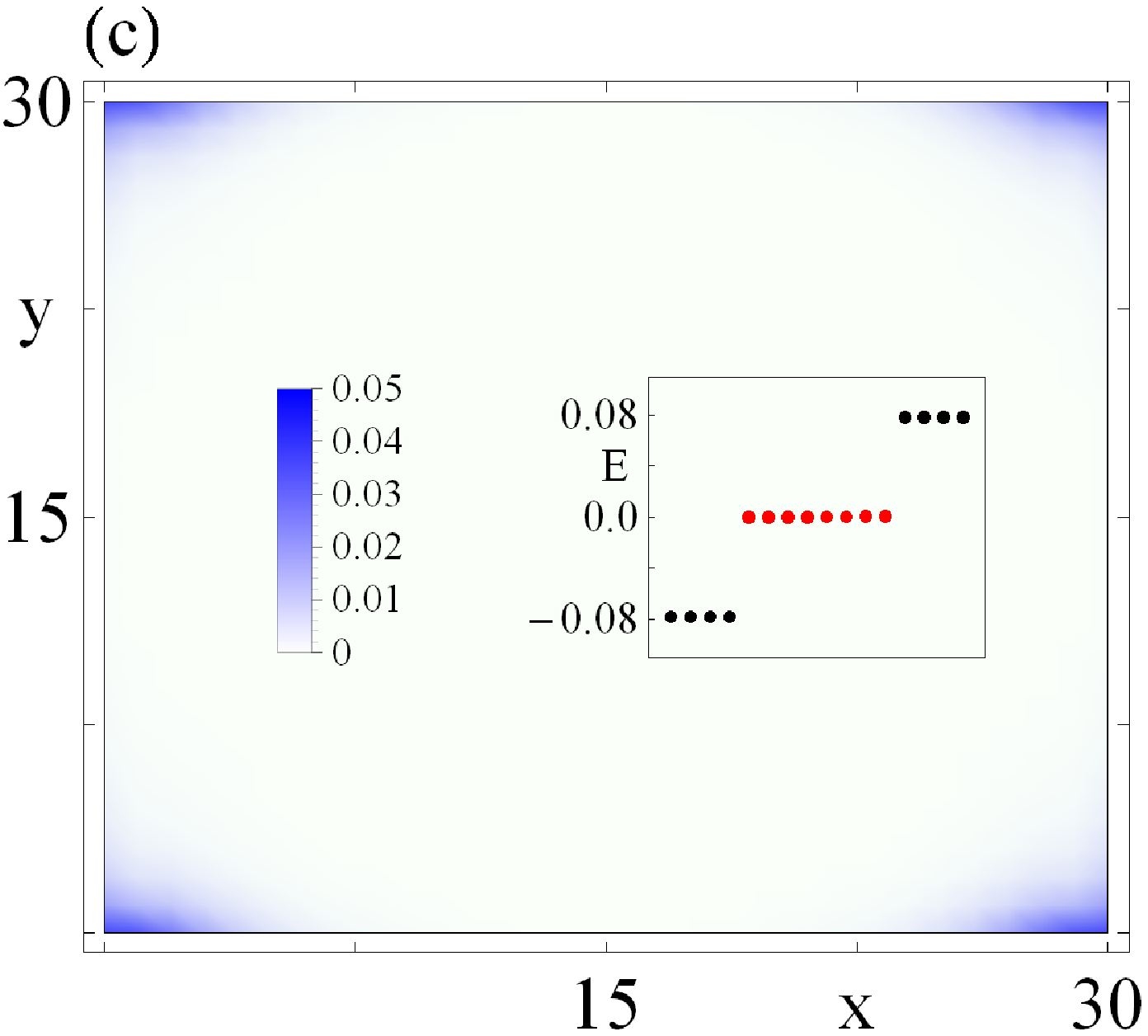}}
\subfigure{\includegraphics[width=4.8cm, height=4.0cm]{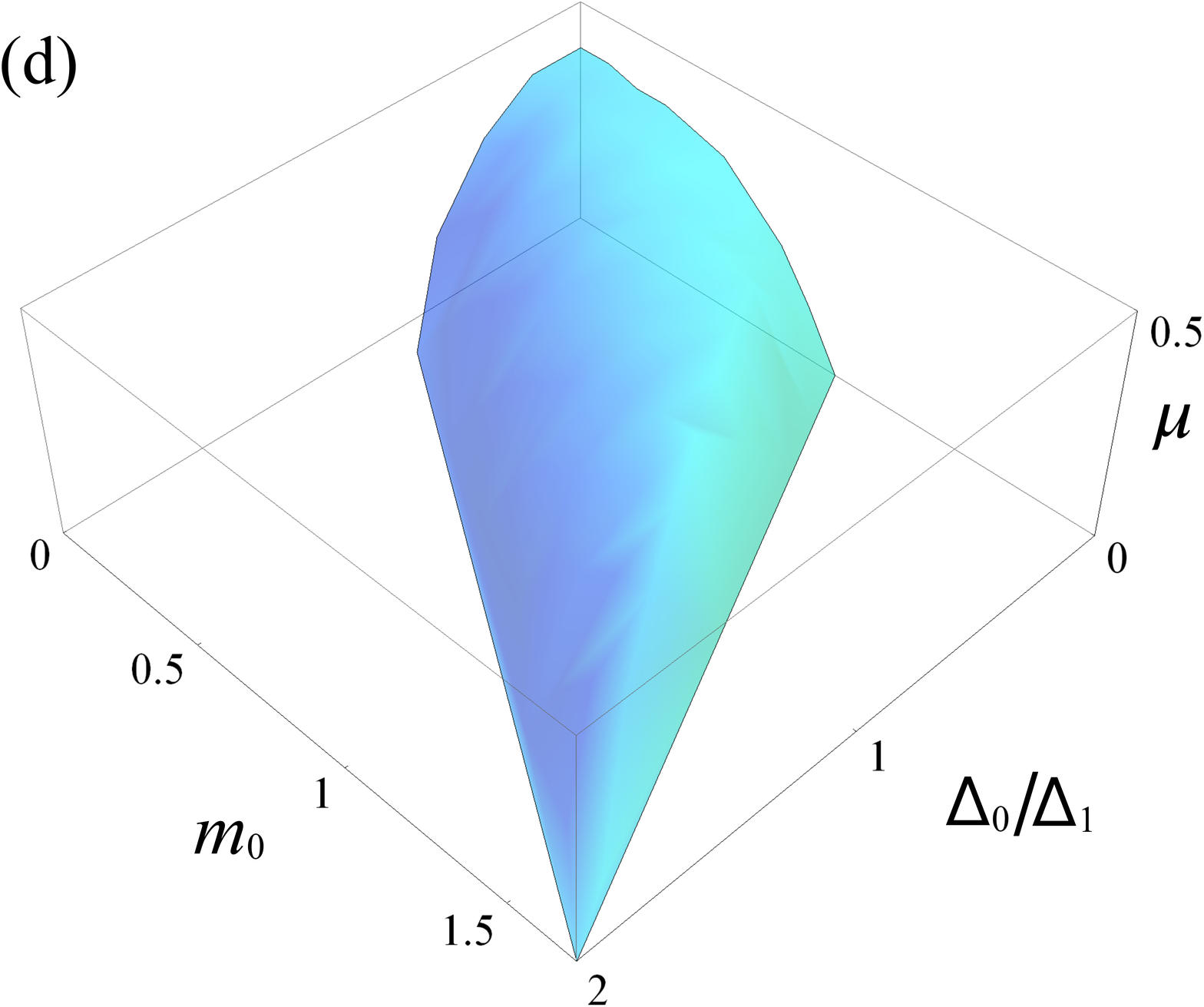}}
\caption{
(a) The pairing nodal ring (red thick line) and band-inversion ring (blue thin line) for $m_{0}=1.0$. The dashed line denotes the Fermi surface for $\mu=0.3$. The wavefunction profiles of MKPs for (b) $\mu=0$ and
(c) $\mu=0.3$, with $\Delta_{0}=\Delta_{1}=0.4$.  (d) $(m_0,\Delta_0/\Delta_1,\mu)$ phase diagram with fixed $\Delta_1=0.4$. Corner modes are found in the surface-enclosed region. Common parameters: $t_{x}=A_{x}=0.4$, $t_{y}=A_{y}=1.3$. }\label{gapped}
\end{figure*}

To have MKP at each corner, the sign of Dirac mass $M(l)$ must change from an edge to its adjacent, which leads to the following criterion:
\begin{eqnarray}
(\bar{\Delta}_{0} +\Delta_{1}m/t_{x})(\bar{\Delta}_{0}+\Delta_{1}m/t_{y})<0. \label{criterion}
\end{eqnarray}
Let us define $R_{s}\equiv\sqrt{2 \bar{\Delta}_{0}/\Delta_{1}}$, whose physical meaning
is the radius of the ring of the pairing node, across which the pairing changes sign, and $R_{x}\equiv\sqrt{-2m/t_{x}}$ and
$R_{y}\equiv\sqrt{-2m/t_{y}}$, whose meanings are the two semi-axes of the ellipse determined by $m+\frac{t_{x}}{2}k_{x}^{2}+\frac{t_{y}}{2}k_{y}^{2}=0$ (i.e., the ``band-inversion ring'' of TI, where the sign of the $\sigma_z$ term changes). The mode existence criterion in Eq.(\ref{criterion})  (for $\mu=0$) becomes
\begin{eqnarray}
(R_{s}-R_{x})(R_{s}-R_{y})<0, \label{crit}
\end{eqnarray}
which means that the band-inversion ring has to cross the pairing nodal ring [Fig.\ref{gapped}(a)]. Although derived from the continuum model, Eq.(\ref{crit}) is quite accurate according to our lattice-model numerical results\cite{supplemental}. Intuitively, the low-energy edge modes come mainly from states near the band inversion (where the bulk states have the lowest energies), and inherit the sign of pairing there. When the two rings cross, the pairing sign at band inversion is opposite at two adjacent edges, which supports MKPs. We emphasize that the TI has to be anisotropic in the $x,y$ directions to satisfy Eq.(\ref{crit}) ($R_x\neq R_y$), which is the case for the high-transition-temperature TI WTe$_2$\cite{Wu2018QSHE}. In Fig.\ref{gapped}(b), one finds the existence of MKP when Eq.(\ref{crit}) is satisfied. Including a modest chemical potential with a Fermi surface is innocuous [Fig.\ref{gapped}(c)], as the Fermi surface can be gapped out by the induced pairing as long as it does not cross the pairing node. A $(m,\Delta_0/\Delta_1,\mu)$ phase diagram is shown in Fig.\ref{gapped}(d).

So far, we have not discussed disorders. We have numerically confirmed that usual disorders such as on-site random potential does not destroy the MKPs\cite{supplemental}. In addition, our proposal does not require atomically precise edges. Modest edge imperfections do not affect the MKPs because they are pinned to zero energy by particle-hole symmetry; ``big'' edge imperfections just create new corners that host their own MKPs, which offer more opportunity to observe MKPs\cite{supplemental}.

Finally, it is useful to mention that, in both the $d$-wave and $s_\pm$-wave cases, a single MZM can be created from the MKP at the corner by killing one mode in the pair. Apparently, TRS must be broken. For example, it can be achieved by adding an in-plane magnetic field with an appropriate magnitude. The physical picture is most transparent in the edge theory (see the Supplemental Material for details\cite{supplemental}).

\emph{Experimental estimations.--}For concreteness, let us focus on the high-temperature $s_\pm$-wave iron-based superconductors. As emphasized above, in the $s_\pm$-wave case the TI band structure is required to be anisotropic in the $x$ and $y$ directions [due to Eq.(\ref{crit})].
Notably, the monolayer WTe$_{2}$, which has recently been confirmed
as a high-temperature TI in experiments\cite{Wu2018QSHE}(up to 100 Kelvin),
has the desired band structure\cite{Qian2014TMD}.
According
to the $\bk\cdot\bp$ model in Ref.\cite{Qian2014TMD}, we fit the parameters to be
$R_{x}=0.41${\AA}$^{-1}$, $R_{y}=0.15${\AA}$^{-1}$ (details are given in the Supplemental Material\cite{supplemental}). The reciprocal
lattice vectors of WTe$_{2}$ along the $x$ and $y$ directions are $G_{x}\simeq 1.0${\AA}$^{-1}$
and $G_{y}\simeq 1.8${\AA}$^{-1}$. Thus, the band-inversion ring reaches close to the Brillouin zone boundary in the $x$ direction, while it stays close to the zone center in the $y$ direction, resembling the advantageous shape of the band-inversion ring in Fig.\ref{gapped}(a). Although an accurate estimation of the magnitude of the induced pairing gap is not available, we note that cuprate superconductors can induce a gap of tens of meV at the surface states of topological insulators\cite{Wang2013proximity,Zareapour2012proximity}; presumably similar order of magnitude can be expected in the present setup. Therefore, among other options, a setup composed of a WTe$_{2}$ monolayer in proximity to a high-$T_c$ iron-based superconductor is promising for the present proposal. A WTe$_{2}$ monolayer in proximity to cuprate superconductors is also promising.

\emph{Conclusions.--}We have shown that a 2D TI with proximity-induced $d$-wave or $s_\pm$-wave pairing, though being topologically trivial as a TRI superconductor, is a promising candidate of high-temperature platform for realizing robust Majorana corner modes. We provide quantitative criteria for this proposal. This Letter may also stimulate further studies of  topologically-trivial-superconductor-based Majorana modes.

{\it Acknowledgements.---}We would like to thank Wei Li for helpful discussions.
This work is supported by NSFC (No. 11674189). Z. Y. is supported in part by the China Postdoctoral Science Foundation (2016M590082).

\emph{Note added:} Recently, there appeared a related preprint\cite{wang2018corner} that focuses on the $s_\pm$-wave case.

\bibliography{dirac}

\clearpage

{\bf Supplemental Material}

\vspace{10mm}

This supplemental material contains eight parts: (i) The derivation of the edge theory for the II, III, IV edges. (ii) The derivation
of the edge theory for the $s_{\pm}$-wave pairing via the lattice model. (iii) Demonstrating
the absence of Majorana Kramers pair when the pairing nodal ring does not cross the band-inversion ring. (iv) Experimental estimations.
(v) Realizing single Majorana zero mode at the corner. (vi) Stability of the Majorana corner modes against disorders.
(vii) Effects of edge imperfections. (viii) Effects of chemical potential and phase diagram.

\section{I. Edge theory of the II, III, IV edges for the $d$-wave pairing}

We start from
the low-energy bulk Hamiltonian around $\bk=(0,0)$. Having $d$-wave pairing in mind, we take $\Delta_{0}=0$.
The low-energy Hamiltonian at $\mu=0$ reads (not imposing any constraint on $\Delta_{x,y}$ at this stage):
\begin{eqnarray}
H(\bk)&=&(m+\frac{t_{x}}{2}k_{x}^{2}+\frac{t_{y}}{2}k_{y}^{2})\sigma_{z}\tau_{z}+A_{x}k_{x}\sigma_{x}s_{z}+A_{y}k_{y}\sigma_{y}\tau_{z}\nonumber\\
&&+[\Delta_{x}+\Delta_{y}-\frac{1}{2}(\Delta_{x}k_{x}^{2}+\Delta_{y}k_{y}^{2})]s_{y}\tau_{y}.
\end{eqnarray}
Note that $m<0$. For the edge III,
the $k_{y}^{2}$ terms can be neglected and the Hamiltonian is decomposed as $H=H_{0}+H_{p}$, with
\begin{eqnarray}
H_{0}(-i\partial_{x},k_{y})&=&(m-t_{x}\partial_{x}^{2}/2)\sigma_{z}\tau_{z}-iA_{x}\partial_{x}\sigma_{x}s_{z},\nonumber\\
H_{p}(-i\partial_{x},k_{y})&=&A_{y} k_{y}\sigma_{y}\tau_{z}+(\Delta_{x}+\Delta_{y}+\frac{\Delta_{x}}{2}\partial_{x}^{2})s_{y}\tau_{y}.
\end{eqnarray}
When solving the eigenvalue equation $H_{0}\psi_{\alpha}(x)=E_{\alpha}\psi_{\alpha}(x)$,
the boundary condition is  $\psi_{\alpha}(0)=\psi_{\alpha}(-\infty)=0$. A straightforward
calculation gives four solutions with $E_{\alpha}=0$, whose forms are
\begin{eqnarray}
\psi_{\alpha}(x)=\mathcal{N}_{x}\sin (\kappa_{1}x) e^{\kappa_{2}x}e^{ik_{y}y}\tilde{\chi}_{\alpha}
\end{eqnarray}
with the normalization constant $\mathcal{N}_{x}=2\sqrt{\kappa_{2}(\kappa_{1}^{2}+\kappa_{2}^{2})/\kappa_{1}^{2}}$
and the two parameters $\kappa_{1}$ and $\kappa_{2}$ given by
\begin{eqnarray}
\kappa_{1}=\sqrt{-\frac{2m}{t_{x}}-\frac{A_{x}^{2}}{t_{x}^{2}}}, \quad\kappa_{2}=\frac{A_{x}}{t_{x}}.
\end{eqnarray}
$\tilde{\chi}_{\alpha}$ are eigenvectors satisfying $\sigma_{y}s_{z}\tau_{z}\tilde{\chi}_{\alpha}=\tilde{\chi}_{\alpha}$.
Here we choose
\begin{eqnarray}
\tilde{\chi}_{1}&=&|\sigma_{y}=+1\rangle\otimes|\uparrow\rangle\otimes|\tau_{z}=+1\rangle,\nonumber\\
\tilde{\chi}_{2}&=&|\sigma_{y}=-1\rangle\otimes|\downarrow\rangle\otimes|\tau_{z}=+1\rangle, \nonumber\\ \tilde{\chi}_{3}&=&|\sigma_{y}=-1\rangle\otimes|\uparrow\rangle\otimes|\tau_{z}=-1\rangle,\nonumber\\
\tilde{\chi}_{4}&=&|\sigma_{y}=+1\rangle\otimes|\downarrow\rangle\otimes|\tau_{z}=-1\rangle.
\end{eqnarray}
Then the matrix elements of the perturbation $H_{p}$ in this basis are
\begin{eqnarray}
H_{{\rm III}, \alpha\beta}(k_{y})=\int_{-\infty}^{0} dx\psi_{\alpha}^{*}(x)H_{p}(-i\partial_{x},k_{y})\psi_{\beta}(x).
\end{eqnarray}
In terms of the Pauli matrices, the final form of the effective Hamiltonian is
\begin{eqnarray}
H_{\rm III}(k_{y})=A_{y} k_{y}s_{z}+M_{\rm III}s_{y}\tau_{y},
\end{eqnarray}
where
\begin{eqnarray}
M_{\rm III}&=&\int_{-\infty}^{0} dx\psi_{\alpha}(x)^{*}(\Delta_{x}+\Delta_{y}+\frac{\Delta_{x}}{2}\partial_{x}^{2})\psi_{\alpha}(x)\nonumber\\
&=&\Delta_{x}+\Delta_{y}+\Delta_{x}m/t_{x}.
\end{eqnarray}

Similarly, for the edge II, we also decompose
the Hamiltonian into two parts, discarding terms of the order of $k_{x}^{2}$:
\begin{eqnarray}
H_{0}(k_{x},-i\partial_{y})&=&(m-t_{y}\partial_{y}^{2}/2)\sigma_{z}\tau_{z}-iA_{y}\partial_{y}\sigma_{y}\tau_{z},\nonumber\\
H_{p}(k_{x},-i\partial_{y})&=&A_{x} k_{x}\sigma_{x}s_{z}+(\Delta_{x}+\Delta_{y}+\frac{\Delta_{y}}{2}\partial_{y}^{2})s_{y}\tau_{y}.
\end{eqnarray}
By solving the eigenvalue equation $H_{0}\psi_{\alpha}(y)=E_{\alpha}\psi_{\alpha}(y)$ with the boundary condition
$\psi_{\alpha}(0)=\psi_{\alpha}(+\infty)=0$, we find that there are four
solutions with $E_{\alpha}=0$, whose forms are
\begin{eqnarray}
\psi_{\alpha}(y)=\mathcal{N}_{y}\sin (\gamma_{1}y) e^{-\gamma_{2}y}e^{ik_{x}x}\xi_{\alpha}
\end{eqnarray}
with the normalization constant $\mathcal{N}_{y}=2\sqrt{\gamma_{2}(\gamma_{1}^{2}+\gamma_{2}^{2})/\gamma_{1}^{2}}$
and the two parameters $\gamma_{1}$ and $\gamma_{2}$ given by
\begin{eqnarray}
\gamma_{1}=\sqrt{-\frac{2m}{t_{y}}-\frac{A_{y}^{2}}{t_{y}^{2}}}, \quad\gamma_{2}=\frac{A_{y}}{t_{y}}.
\end{eqnarray}
$\xi_{\alpha}$ are the eigenvectors satisfying $\sigma_{x}\xi_{\alpha}=\xi_{\alpha}$.
Here we choose
\begin{eqnarray}
\xi_{1}&=&|\sigma_{x}=+1\rangle\otimes|\uparrow\rangle\otimes|\tau_{z}=+1\rangle,\nonumber\\
\xi_{2}&=&|\sigma_{x}=+1\rangle\otimes|\downarrow\rangle\otimes|\tau_{z}=+1\rangle, \nonumber\\
\xi_{3}&=&|\sigma_{x}=+1\rangle\otimes|\uparrow\rangle\otimes|\tau_{z}=-1\rangle,\nonumber\\
\xi_{4}&=&|\sigma_{x}=+1\rangle\otimes|\downarrow\rangle\otimes|\tau_{z}=-1\rangle.
\end{eqnarray}
In this basis, the matrix elements of the perturbation $H_{p}$ are
\begin{eqnarray}
H_{{\rm II}, \alpha\beta}(k_{x})=\int_{0}^{+\infty} dy\psi_{\alpha}^{*}(y)H_{p}(k_{x},-i\partial_{y})\psi_{\alpha}(y).
\end{eqnarray}
In terms of the Pauli matrices, the final form of the effective Hamiltonian is
\begin{eqnarray}
H_{\rm II}(k_{x})=A_{x} k_{x}s_{z}+M_{\rm II}s_{y}\tau_{y},
\end{eqnarray}
where
\begin{eqnarray}
M_{\rm II}&=&\int_{0}^{+\infty} dy\psi_{\alpha}^{*}(y)(\Delta_{x}+\Delta_{y}+\frac{\Delta_{y}}{2}\partial_{y}^{2})\psi_{\alpha}(y)\nonumber\\
&=&\Delta_{x}+\Delta_{y}+\Delta_{y}m/t_{y}.
\end{eqnarray}
Similarly, for the edge IV, the effective Hamiltonian
is
\begin{eqnarray}
H_{\rm IV}(k_{x})=-A_{x} k_{x}s_{z}+M_{\rm IV}s_{y}\tau_{y},
\end{eqnarray}
and $M_{\rm IV}=M_{\rm II}$.

For the $d$-wave pairing with amplitude satisfying $\Delta_{x}=-\Delta_{y}$,
the $\Delta_{x}+\Delta_{y}$ term appearing in the mass term vanishes.
Let $\Delta_{x}=-\Delta_{y}\equiv\Delta_{d}$, then the effective Hamiltonian of
the four edges are
\begin{eqnarray}
H_{\rm I}(k_{y})&=&-A_{y}k_{y}s_{z}+M_{\rm I}s_{y}\tau_{y},\nonumber\\
H_{\rm II}(k_{x})&=&A_{x}k_{x}s_{z}+M_{\rm II}s_{y}\tau_{y},\nonumber\\
H_{\rm III}(k_{y})&=&A_{y}k_{y}s_{z}+M_{\rm III}s_{y}\tau_{y},\nonumber\\
H_{\rm IV}(k_{x})&=&-A_{x}k_{x}s_{z}+M_{\rm IV}s_{y}\tau_{y},
\end{eqnarray}
where $M_{\rm I}=M_{\rm III}=\Delta_{d}m/t_{x}$, $M_{\rm II}=M_{\rm IV}=-\Delta_{d}m/t_{y}$.
It is immediately clear that the mass terms on two neighboring edges always have opposite
signs.

\section{II. Edge theory for the $s_{\pm}$-wave pairing via solving
the lattice model}

We start from the Bogoliubov-de Gennes Hamiltonian $H=\sum_{k}\Psi_{k}^{\dag}H(\bk)\Psi_{k}$
with $\Psi_{k}=(c_{a,k\uparrow},c_{b,k\uparrow}, c_{a,k\downarrow},c_{b,k\downarrow},
c_{a,-k\uparrow}^{\dag},c_{b,-k\uparrow}^{\dag}, c_{a,-k\downarrow}^{\dag},c_{b,-k\downarrow}^{\dag})^{T}$ and
\bea
H({\bf k})=(m_0-t_x\cos k_x-t_y\cos k_y)\Gamma_{1}+A_x \sin k_x\Gamma_{2}\nonumber\\+A_y \sin k_y\Gamma_{3}
+[\Delta_0-\Delta_{1}(\cos k_x+\cos k_y)]\Gamma_{4}. \eea
where $\Gamma_{1}=\sigma_{z}\tau_{z}$, $\Gamma_{2}=\sigma_{x}s_{z}$,
$\Gamma_{3}=\sigma_{y}\tau_{z}$ and $\Gamma_{4}=s_{y}\tau_{y}$. Here we have taken the chemical potential $\mu=0$.

We first investigate the edge I. As the system takes open boundary
condition in the $x$ direction and periodic boundary condition in the $y$ direction, we do
a partial Fourier transformation of the Hamiltonian, which gives
\begin{widetext}
\begin{eqnarray}
H&=&\sum_{x,k_{y}}\Psi_{x,k_{y}}^{\dag}\left[(m_{0}-t_{y}\cos k_{y})\Gamma_{1}+A_{y}\sin k_{y}\Gamma_{3}+
(\Delta_0-\Delta_{1}\cos k_y)\Gamma_{4}\right]\Psi_{x,k_{y}}\nonumber\\
&&+\sum_{x,k_{y}}\left[\Psi_{x,k_{y}}^{\dag}(-\frac{t_{x}}{2}\Gamma_{1}+i\frac{A_{x}}{2}\Gamma_{2}-\frac{\Delta_{1}}{2}\Gamma_{4})
\Psi_{x+1,k_{y}}+h.c.\right], \text{with}\nonumber\\
\Psi_{x,k_{y}}&=&(c_{x; a,k_{y}\uparrow},c_{x; b,k_{y}\uparrow}, c_{x; a,k_{y}\downarrow},c_{x; b,k_{y}\downarrow},
c_{x; a,-k_{y}\uparrow}^{\dag},c_{x; b,-k_{y}\uparrow}^{\dag}, c_{x; a,-k_{y}\downarrow}^{\dag},c_{x; b,-k_{y}\downarrow}^{\dag})^{T}.
\end{eqnarray}
\end{widetext}
where $x$ is the integer-valued coordinate (the lattice constant $a=1$) taking values from $1$ to $L$. In the basis $(\Psi_{1,k_{y}},\Psi_{2,k_{y}},...)$, the Hamiltonian
can be expressed in a matrix form
$H(k_{y})=H_0(k_y)+H_1(k_y)+H_{2}(k_y)$ with
\bea
H_0(k_y)=
\begin{pmatrix}
D_0&T_0&0&\cdots \\T_0^{\dag}&D_0&T_{0}&\cdots \\0&T_0^{\dag}&D_0&\cdots \\\vdots &\vdots &\vdots &\ddots
\end{pmatrix},
\eea
where $D_0=(m_0-t_y\cos k_y), T_0= (-t_x\Gamma_1+i A_x\Gamma_2)/2$,
\bea
H_1(k_y)=
\begin{pmatrix}
D_1&T_1&0&\cdots \\T_1^{\dag}&D_1&T_{1}&\cdots \\0&T_1^{\dag}&D_1&\cdots \\\vdots &\vdots &\vdots &\ddots
 \end{pmatrix},
 \eea
where $D_1=(\Delta_0-\Delta_1 \cos k_y)\Gamma_4, T_1= -\Delta_1\Gamma_4/2$,
and
\bea
H_2(k_y)=
\begin{pmatrix}
A_y \sin k_y\Gamma_3&0&0&\cdots \\0&A_y \sin k_y\Gamma_3 &0&\cdots \\0&0&A_y \sin k_y\Gamma_3&\cdots \\\vdots &\vdots &\vdots &\ddots
 \end{pmatrix}.
 \eea
To simply the calculation, we take $k_y$ to be close to $0$ and the pairing amplitude to be small,
so that $ H_1(k_y)$ and $H_{2}(k_y)$ can be treated as perturbations. We first solve for the eigenstates of $H_0(k_y=0)$.
Their forms can be written as $\psi=(\psi_1,\psi_2\cdots )^{T}$, which satisfies the following iteration
relation,
\bea
\frac{1}{2}(-t\Gamma_1+i A\Gamma_2)\psi_{n+1}+m\Gamma_1\psi_n+\frac{1}{2}(-t\Gamma_1-i A\Gamma_2)\psi_{n-1}=0,\qquad\label{iteration}
\eea
here for convenience we have renamed $m_0-t_y$, $t_x$, $A_x$ as $m$, $t$, $A$, respectively (Do not confuse it with the $m$ in the main text).
Because of the anticommutation relation between $\Gamma_1$ and $\Gamma_2$, the eigenvalues of
$i\Gamma_1\Gamma_2$ are $\pm 1$. To solve this equation, we choose a trial solution $\psi_{\alpha;n}=\lambda^n \phi_\alpha$,
where $\phi_\alpha$ satisfies $-i\Gamma_1\Gamma_2\phi_\alpha=s\phi_{\alpha}$ ($s=\pm 1$).
Note that $-i\Gamma_{1}\Gamma_{2}$ is an $8\times8$ matrix, therefore, there are four $\phi_{\alpha}$'s
satisfying this equation.
Substituting this trial solution into Eq.(\ref{iteration}),
it is readily found that
\bea
\lambda_{\pm}=\frac{m\pm\sqrt{m^2-(t^2-A^2)}}{t+As}.
\eea
It can be shown that $ |\lambda_{\pm}|<1$ when $s=\text{sgn}(t/A)$ and $m^2<t^2$.
If we consider a semi-infinite geometry with $L\rw\infty$, the wave function should satisfy the two boundary
conditions: $\psi_{\alpha;0}=0$ and $\psi_{\alpha;+\infty}=0$ (For convenience, we have added an artificial site $x=0$, on which the wavefunction is zero). The solution is of the form
$\psi_{\alpha;n}=N(\lambda_{+}^n-\lambda_{-}^n)$$\phi_{\alpha}$,  where $N$ is the normalization constant:
\bea
|N|^2=\left[\frac{ts}{A}\frac{|m^2-(t^2-A^2)|}{t^2-m^2}\right]^{-1}.
\eea
Now the effective Hamiltonian for the edge states is obtained from
perturbation theory,
\begin{eqnarray}
H_{{\rm I};\alpha\beta}(k_{y})=\psi_{\alpha}^{+}[H_{1}(k_{y})+H_{2}(k_{y})]\psi_{\beta}.
\end{eqnarray}

The mass term of the effective Hamiltonian comes solely from $H_{1}(k_y)$.
Ignoring all terms of orders higher than $k_{y}$,  we get
\bea
M_{\rm I}=\Delta_0-\Delta_1-\frac{\Delta_1}{2}(\sum_{n=1}\psi_{n+1}^{\dag}\psi_{n}+\sum_{n=2}\psi_{n-1}^{\dag}\psi_{n}).\nonumber
\eea
After straightforward calculations, we find
\bea
M_{\rm I}=\Delta_0-\Delta_1-\Delta_1 |N|^2\frac{ms}{A}\frac{|m^2-(t^2-A^2)|}{t^2-m^2} \nonumber\\
=\Delta_0-\Delta_1-\Delta_1\frac{m_0-t_y}{t_x}.
\eea
Similar calculations lead to the mass terms for the other three edges, II, III, IV, which
are $M_{\rm III}=M_{\rm I}$ and
\begin{eqnarray}
M_{\rm II}=M_{\rm IV}
=\Delta_0-\Delta_1-\Delta_1\frac{m_0-t_x}{t_y}.
\end{eqnarray}
To create corner Majorana Kramers pairs, the mass term must change sign  at the corner,  which requires
\bea
(\Delta_0-\Delta_1-\Delta_1\frac{m_0-t_x}{t_y})(\Delta_0-\Delta_1-\Delta_1\frac{m_0-t_y}{t_x})<0.
\eea
This criterion is the same as the one obtained from continuum model in the main text.

\begin{figure*}[t!]
\subfigure{\includegraphics[width=5cm, height=5cm]{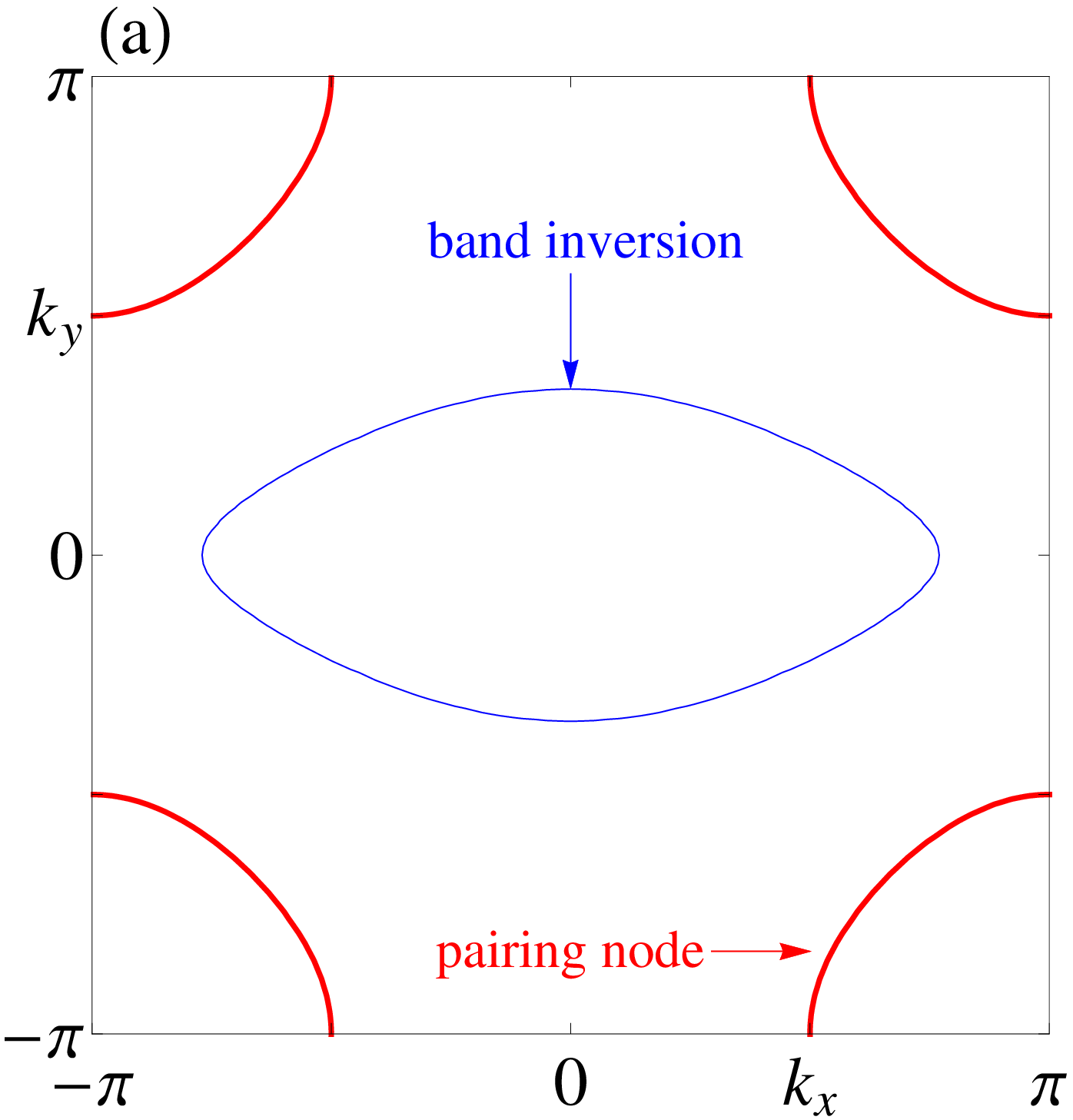}}
\subfigure{\includegraphics[width=5cm, height=5cm]{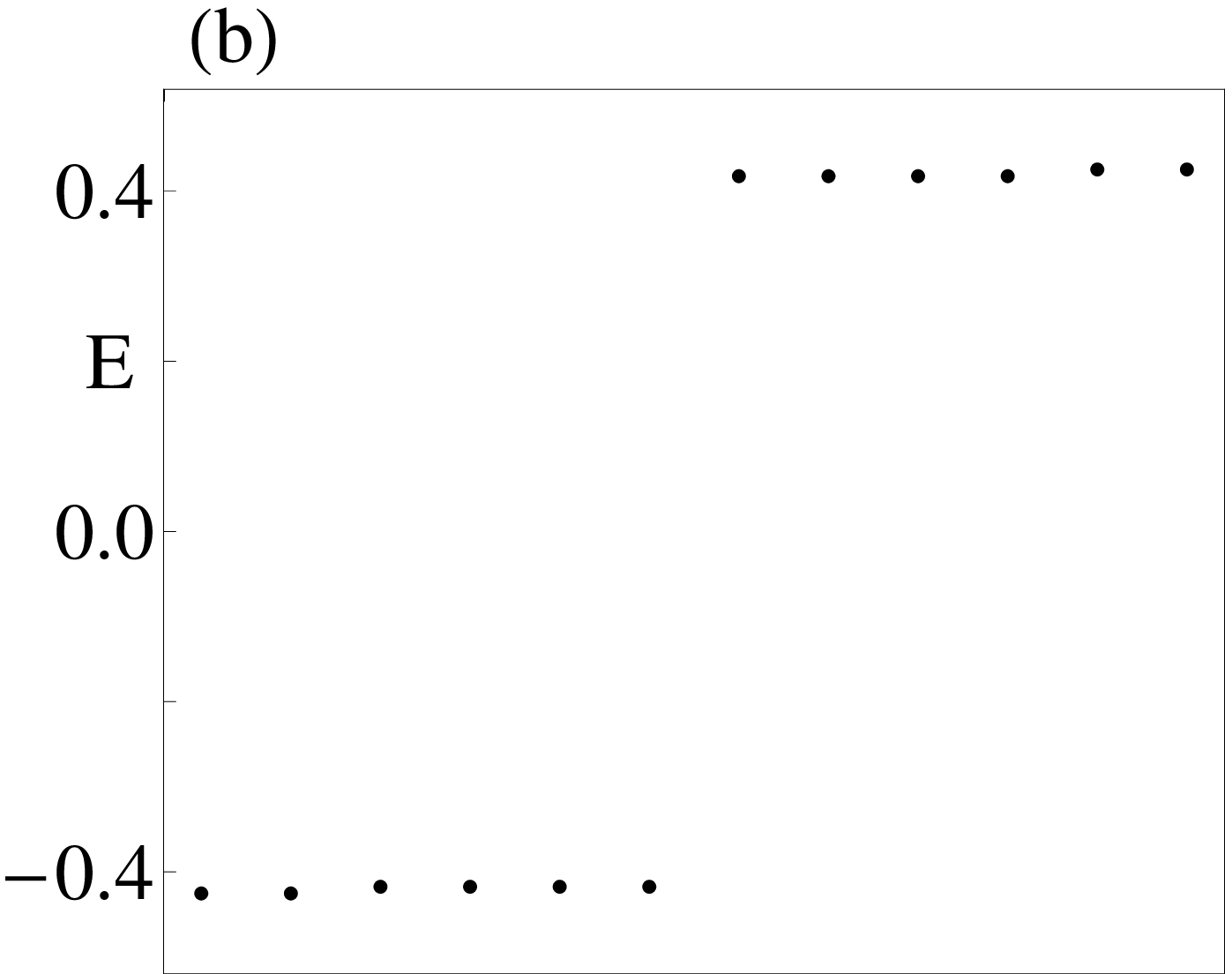}}
\caption{When the pairing nodal ring does not cross the band-inversion ring, the Majorana Kramers pair is found to be absent. The parameters are $m_{0}=1.0$, $t_{x}=A_{x}=0.4$, $t_{y}=A_{y}=1.3$, $\Delta_{0}=0.4$, $\Delta_{1}=-0.4$, $\mu=0$,
$L_{x}=L_{y}=30$.}\label{absent}
\end{figure*}

\section{III. Majorana Kramers pair is absent when the pairing nodal ring does not cross the
band-inversion ring}

In the main text, we have shown when the pairing nodal ring crosses the
band-inversion ring, Majorana Kramers pairs are created at the corner of TI.
To display the opposite situation, we tune the pairing nodal ring of Fig.4(a) in the main text to be around the $(\pi,\pi)$ point while keeping the
band-inversion ring unchanged, so that the two rings no longer cross each other,
as shown in Fig.\ref{absent}(a).
In this regime,  the numerical results demonstrate that there is
no zero energy state (see Fig.\ref{absent}(b)), indicating
the absence of Majorana Kramers pair.

\section{IV. Details of experimental estimations}

In this section, we give an experimental estimation based on the band structures
of WTe$_{2}$, which has recently been found as a 2D TI at temperature as high as 100 Kelvin\cite{Wu2018QSHE}. In Ref.\cite{Qian2014TMD}, a $\bk\cdot\bp$ model has been obtained
to fit the band structure near the $\Gamma$ point (the band-inverted region), which is
\begin{widetext}
\begin{eqnarray}
H(\bk)=\left(
         \begin{array}{cccc}
           -\delta-\frac{\hbar^{2}k_{x}^{2}}{2m_{x}^{p}}-\frac{\hbar^{2}k_{y}^{2}}{2m_{y}^{p}} & 0 & -i\hbar v_{1}k_{x} & \hbar v_{2}k_{y} \\
           0 & -\delta-\frac{\hbar^{2}k_{x}^{2}}{2m_{x}^{p}}-\frac{\hbar^{2}k_{y}^{2}}{2m_{y}^{p}} & \hbar v_{2}k_{y} & -i\hbar v_{1}k_{x} \\
           i\hbar v_{1}k_{x} & \hbar v_{2}k_{y} & \delta+\frac{\hbar^{2}k_{x}^{2}}{2m_{x}^{d}}+\frac{\hbar^{2}k_{y}^{2}}{2m_{y}^{d}} & 0 \\
           \hbar v_{2}k_{y} & i\hbar v_{1}k_{x} & 0 & \delta+\frac{\hbar^{2}k_{x}^{2}}{2m_{x}^{d}}+\frac{\hbar^{2}k_{y}^{2}}{2m_{y}^{d}} \\
         \end{array}
       \right),
\end{eqnarray}
\end{widetext}
where $v_{1}=3.87\times 10^{5}$ m/s, $v_{2}=0.46\times 10^{5}$ m/s, $\delta=-0.33$ eV, $m_{x}^{p}=0.50m_{e}$,
$m_{y}^{p}=0.16m_{e}$, $m_{x}^{d}=2.48m_{e}$, $m_{y}^{d}=0.37m_{e}$, $m_{e}$ being the free electron mass. The lattice
constant of WTe$_{2}$ is $a=6.25${\AA}, $b=3.48${\AA}. To simply the calculation, we make an approximation that $m_{x}=\sqrt{m_{x}^{p}m_{x}^{d}}=1.11m_{e}$
and $m_{y}=\sqrt{m_{y}^{p}m_{y}^{d}}=0.24m_{e}$, and then transform the $\bk\cdot\bp$ model to the lattice form,
which is
\begin{eqnarray}
H(\bk)&=&[m_{0}+t_{x}\cos (k_{x}a)+t_{y}\cos (k_{y}b)]\sigma_{z}+A_{x}\sin (k_{x}a)\sigma_{y}\nonumber\\
&&+A_{y}\sin (k_{y}b)s_{x}\sigma_{x},
\end{eqnarray}
where $t_{x}=\frac{\hbar^{2}}{m_{x}a^{2}}=0.18$ eV, $t_{y}=\frac{\hbar^{2}}{m_{y}b^{2}}=2.64$ eV,
$m_{0}=-\delta-t_{x}-t_{y}=-2.49$ eV.
$A_{x}=\hbar v_{1}/a=0.41$ eV, $A_{y}=\hbar v_{2}/b=0.09$ eV.
The band-inversion ring intersects the $k_{x}$ axis at $R_{x}=\arccos[-(m_{0}+t_{y})/t_{x}]/a=0.41${\AA}$^{-1}$, and the $y$ axis at
$R_{y}=\arccos[-(m_{0}+t_{x})/t_{y}]/b=0.15${\AA}$^{-1}$. The energy gap is $0.087$ eV (at $(k_{x},k_{y})=(0,\pm R_{y})$). It is notable that $R_{y}$ and the energy gap thus obtained agree excellently with
the results ($0.146${\AA}$^{-1}$, about $0.08$ eV) based on the DFT calculation\cite{Qian2014TMD}, indicating this lattice model gives an accurate
description of the relevant band structure. In addition, the reciprocal lattice vectors are
$G_{x}=2\pi/a=1.00${\AA}$^{-1}$, $G_{y}=2\pi/b=1.80${\AA}$^{-1}$, and it is straightforward to find that $R_{x}/(G_{x}/2)=0.82$,
$R_{y}/(G_{y}/2)=0.16$, indicating that the band-inversion ring reaches close to the Brillouin-zone boundary in the $x$ direction, while stays close to the zone center in the $y$ direction.

\begin{figure*}[t!]
\subfigure{\includegraphics[width=5cm, height=5cm]{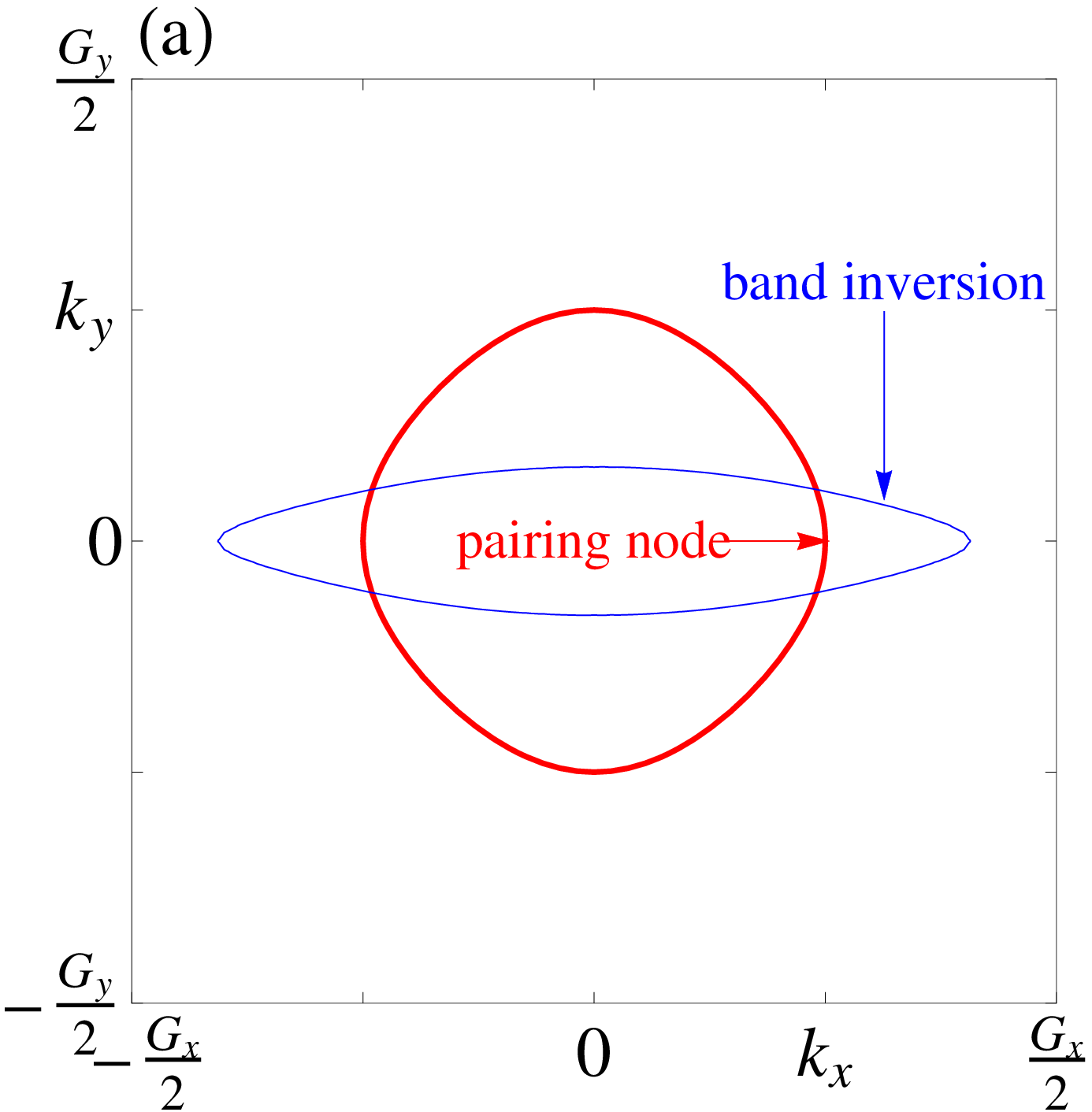}}
\subfigure{\includegraphics[width=5cm, height=5cm]{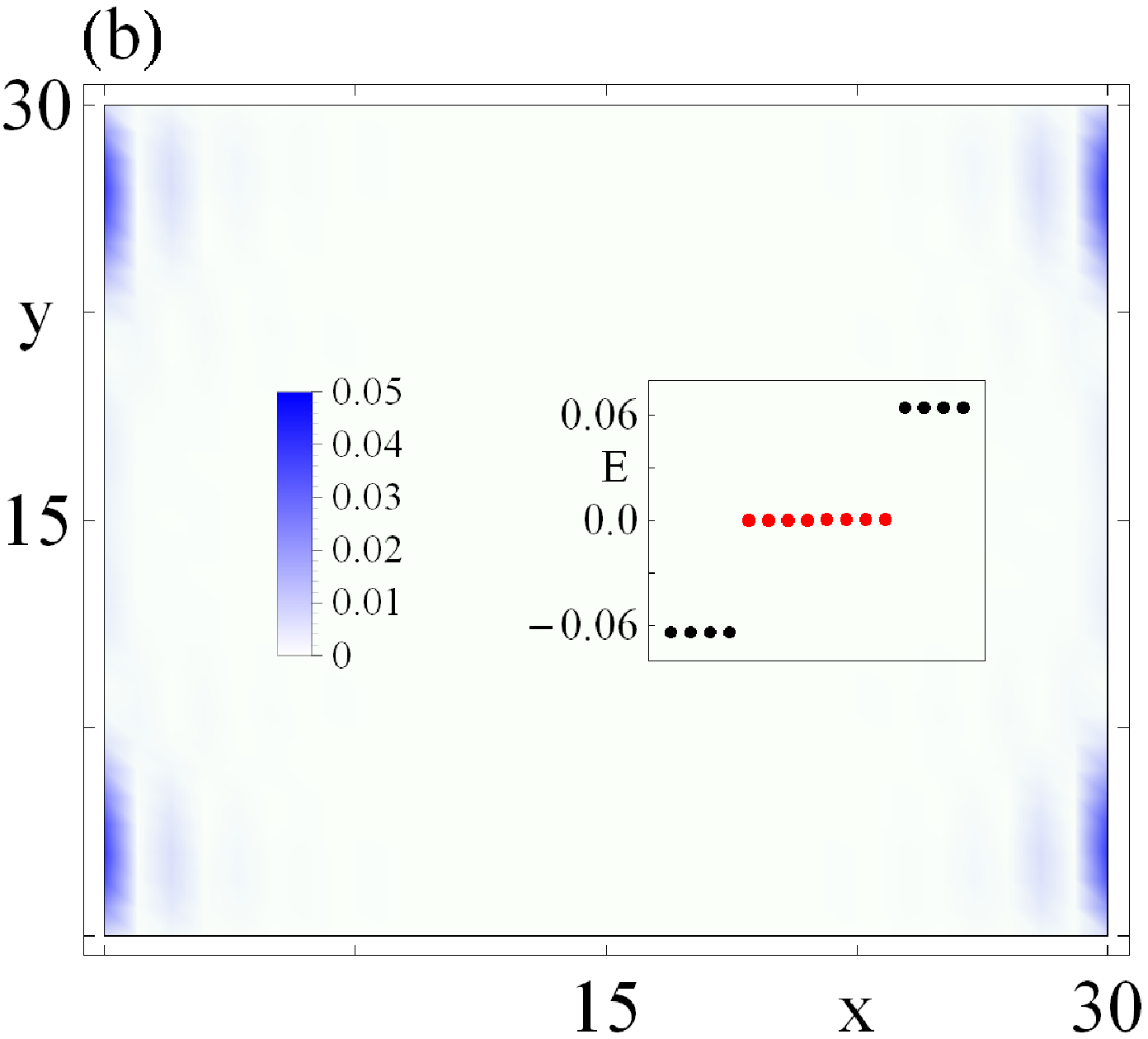}}
\caption{
(a) The thick red line is the nodal ring of the $s_{\pm}$-wave pairing, and the
thin blue line is the band-inversion ring. (b) The wavefunction profile of the Majorana Kramers pairs
in a $L_{x}\times L_{y}=30\times30$ square sample. The inset
shows energies close to zero ( in units of eV).   $m_{0}=-2.49$ eV, $t_{x}=0.18$ eV, $t_{y}=2.64$ eV, $A_{x}=0.41$ eV, $A_{y}=0.2$ eV, $\Delta_{0}=\Delta_{1}=0.1$ eV. The chemical potential is taken to be zero.  }\label{experiment}
\end{figure*}

In the presence of pairing, the Hamiltonian becomes
\begin{eqnarray}
H(\bk)&=&[m_{0}+t_{x}\cos (k_{x}a)+t_{y}\cos (k_{y}b)]\sigma_{z}\tau_{z}+A_{x}\sin (k_{x}a)\sigma_{y}\tau_{z}\nonumber\\
&&+A_{y}\sin (k_{y}b)s_{x}\sigma_{x}+\Delta(\bk)s_{y}\tau_{y},
\end{eqnarray}
where $\Delta(\bk)=\Delta_{0}+\Delta_{x}\cos (k_{x}a)+\Delta_{y}\cos (k_{y}b)$.
The only difference between this Hamiltonian
and the one in Eq.(1) of the main text is in the basis choices.

In high-temperature cuprate superconductor, the reduced gap $2\Delta/k_{B}T$
can be much larger than the expected BCS value $4.3$ for $d$-wave pairing\cite{Won1994}, indicating that the pairing amplitude
can be quite large. e.g., it was found that $\Delta$ can be as high as $60$ meV in
superconductor Bi$_{2}$Sr$_{2}$Ca$_{2}$Cu$_{3}$O$_{10+\delta}$ ($T_{c}=109$ K)\cite{Kugler2006Bi2223}. In high-temperature
iron-based superconductor, the pairing amplitude can also be higher than $10$ meV, e.g.,
$\Delta=15$ meV in superconductor Ba$_{1-x}$K$_{x}$Fe$_{2}$As$_{2}$ ($T_{c}=37$ K)\cite{Hoffman2011review}.

In Fig.\ref{experiment}, we use the model parameters of WTe$_{2}$  extracted from
the $\bk\cdot\bp$ model, and take the $s_\pm$-wave pairing with an amplitude
$\sim 100$ meV (which has been exaggerated, yet the result is qualitatively unchanged; similarly, the value of $A_{y}$ has also been increased  from $0.09$ eV to $0.2$ eV).
Fig.\ref{experiment} demonstrates that Majorana Kramers pairs are created at the corners when the pairing nodal ring crosses
the band-inversion ring.

\section{V. Realizing single Majorana zero mode at the corner}

\begin{figure*}[thb]
\subfigure{\includegraphics[width=3.75cm, height=4cm]{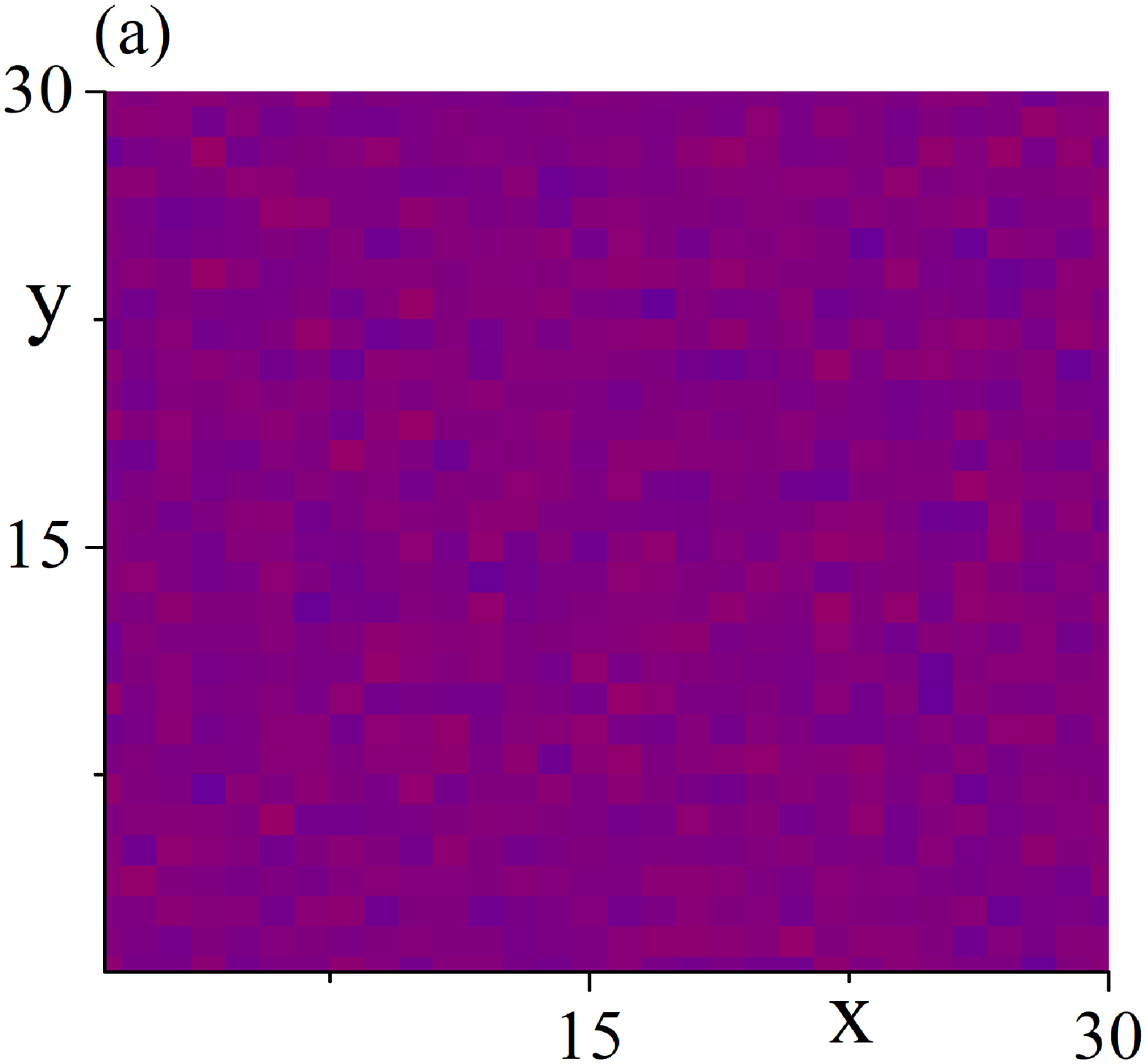}}
\subfigure{\includegraphics[width=3.75cm, height=4cm]{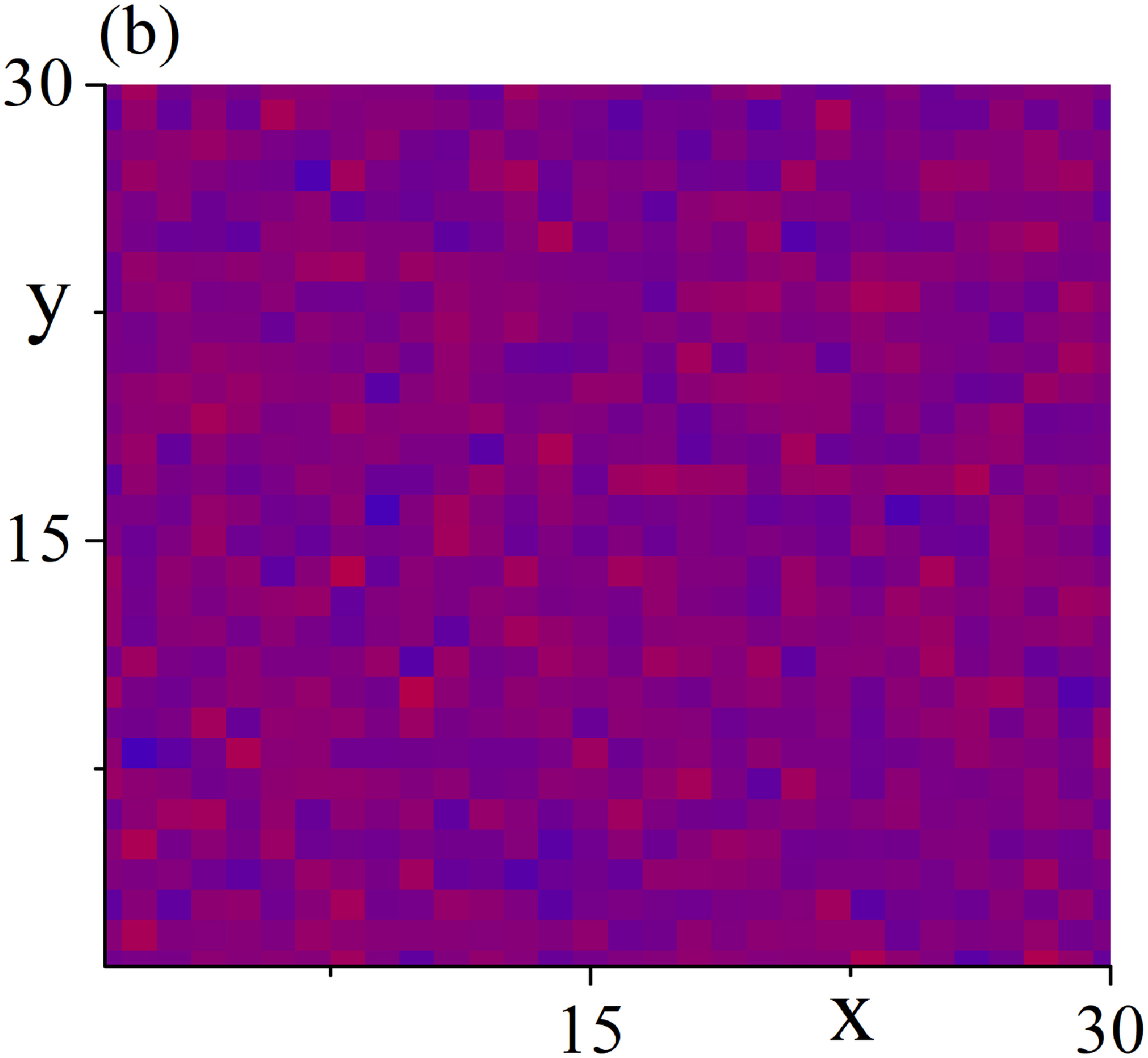}}
\subfigure{\includegraphics[width=3.75cm, height=4cm]{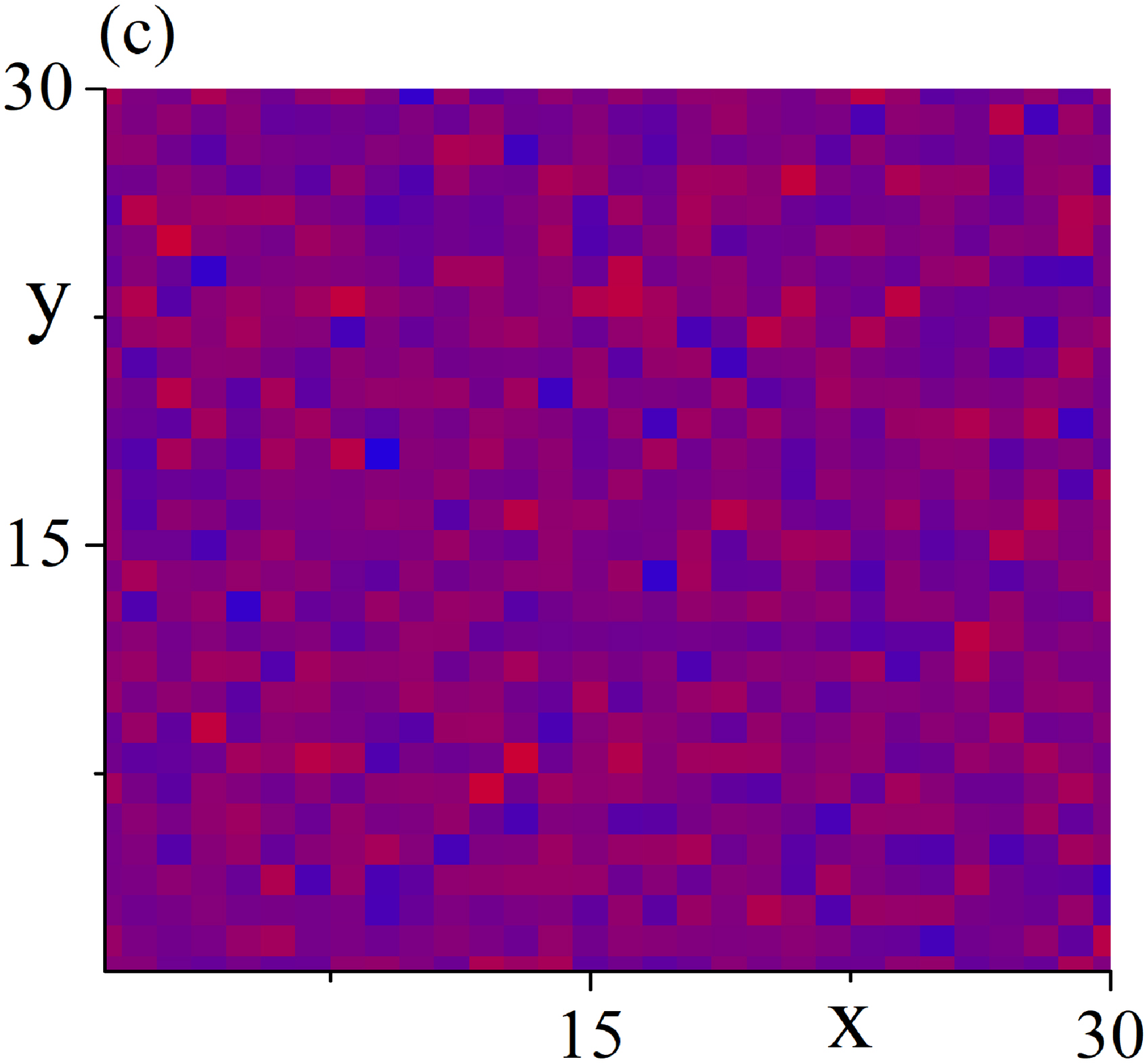}}
\subfigure{\includegraphics[width=4.75cm, height=4cm]{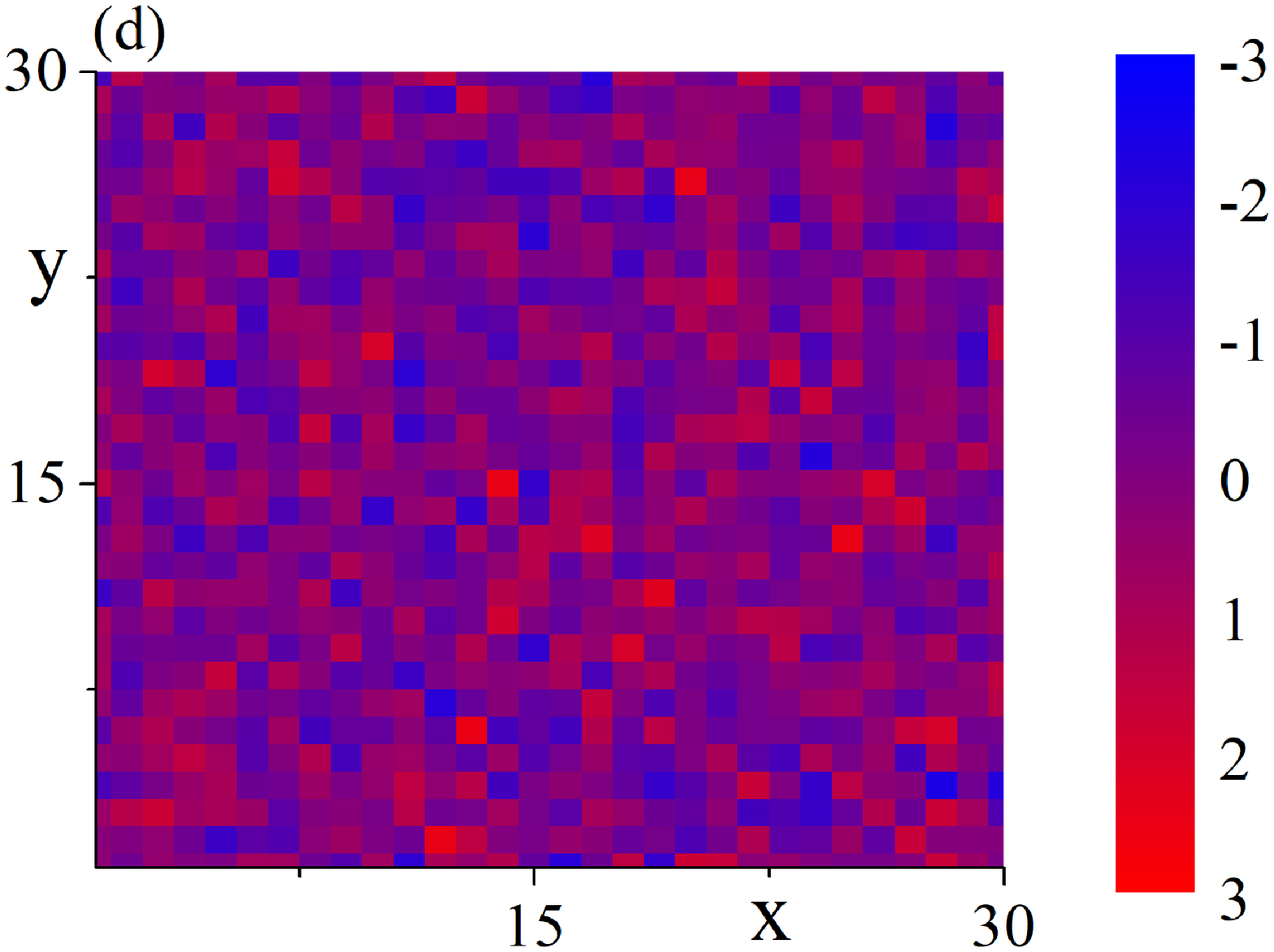}}
\subfigure{\includegraphics[width=3.75cm, height=4cm]{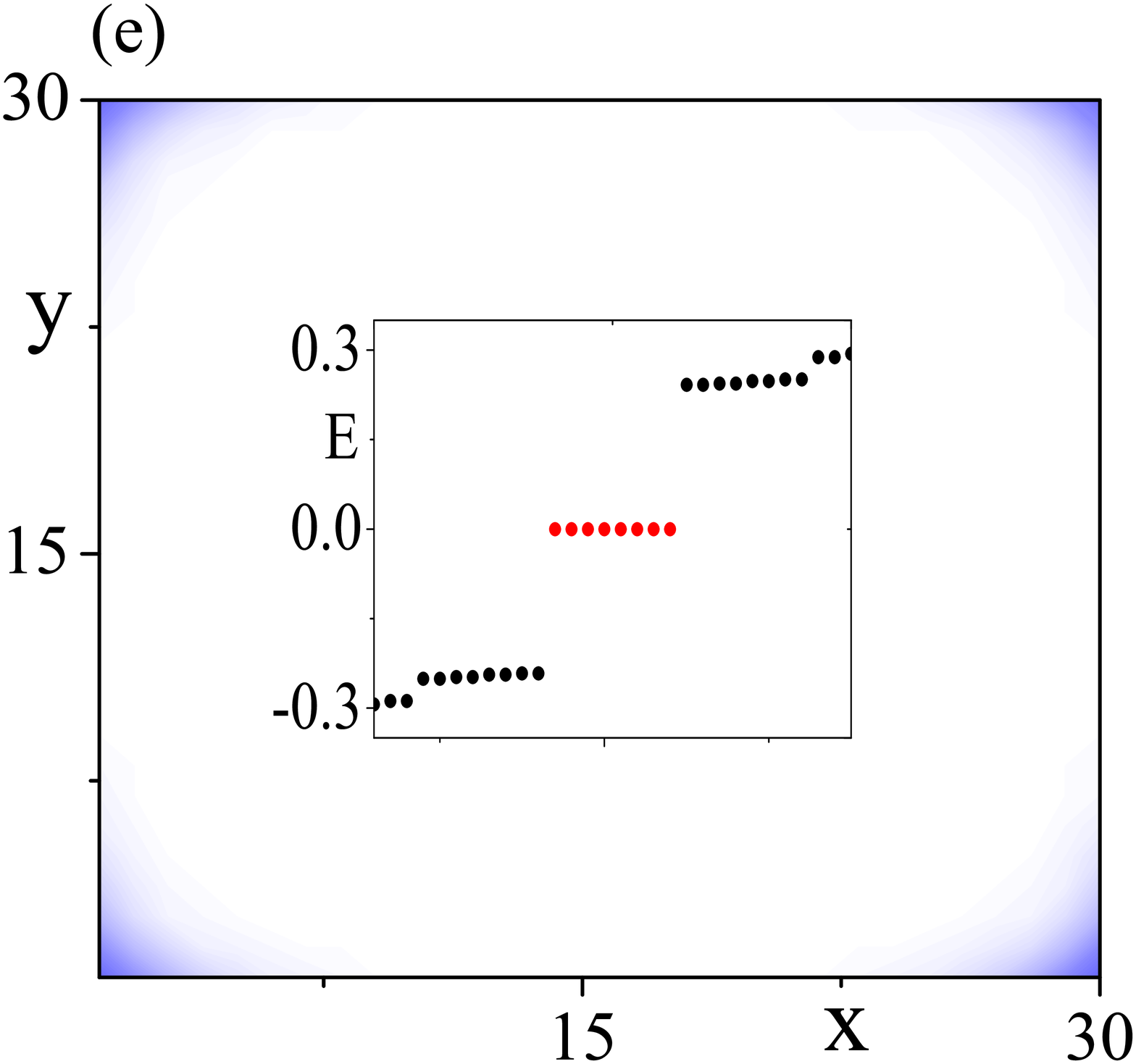}}
\subfigure{\includegraphics[width=3.75cm, height=4cm]{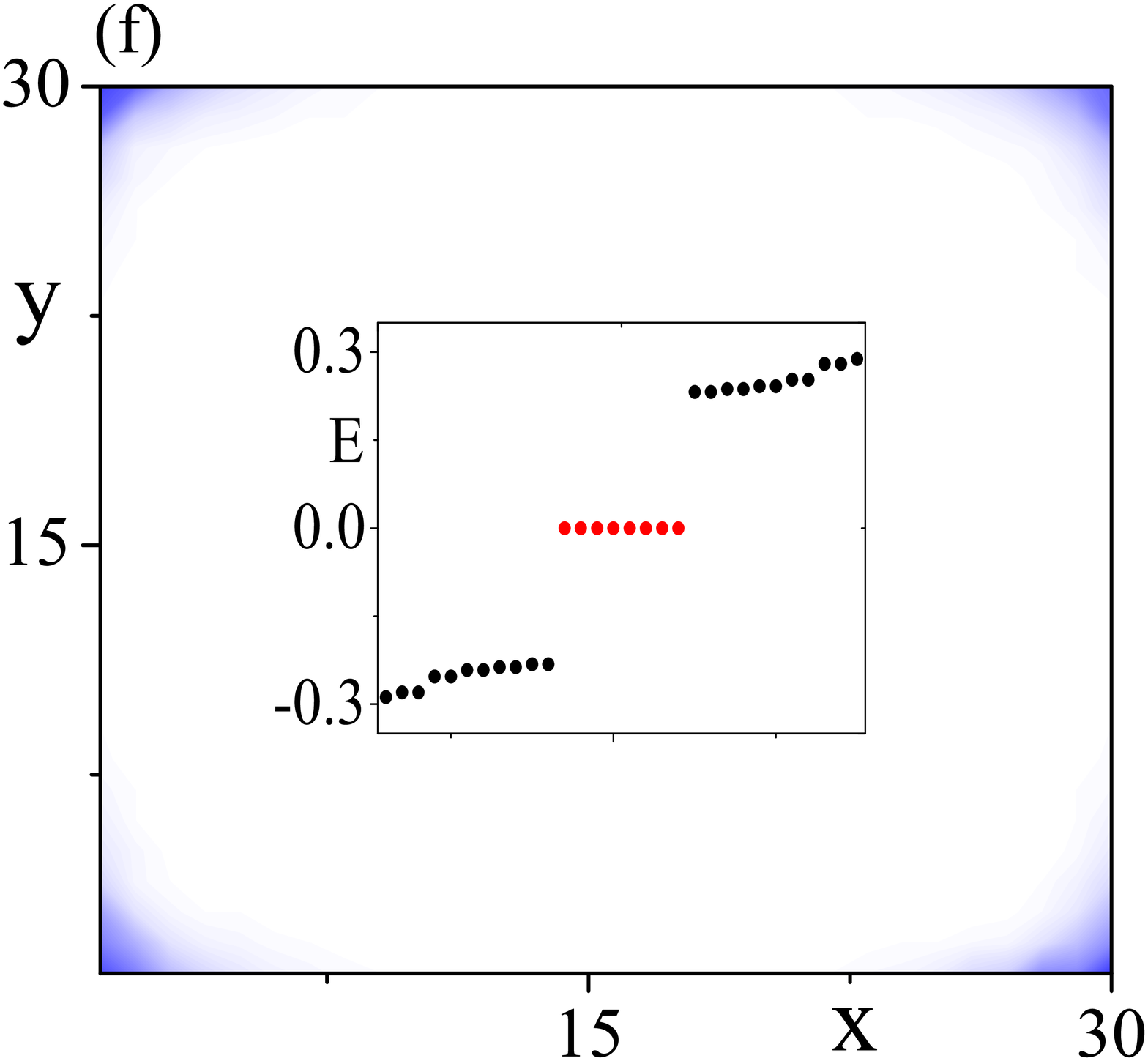}}
\subfigure{\includegraphics[width=3.75cm, height=4cm]{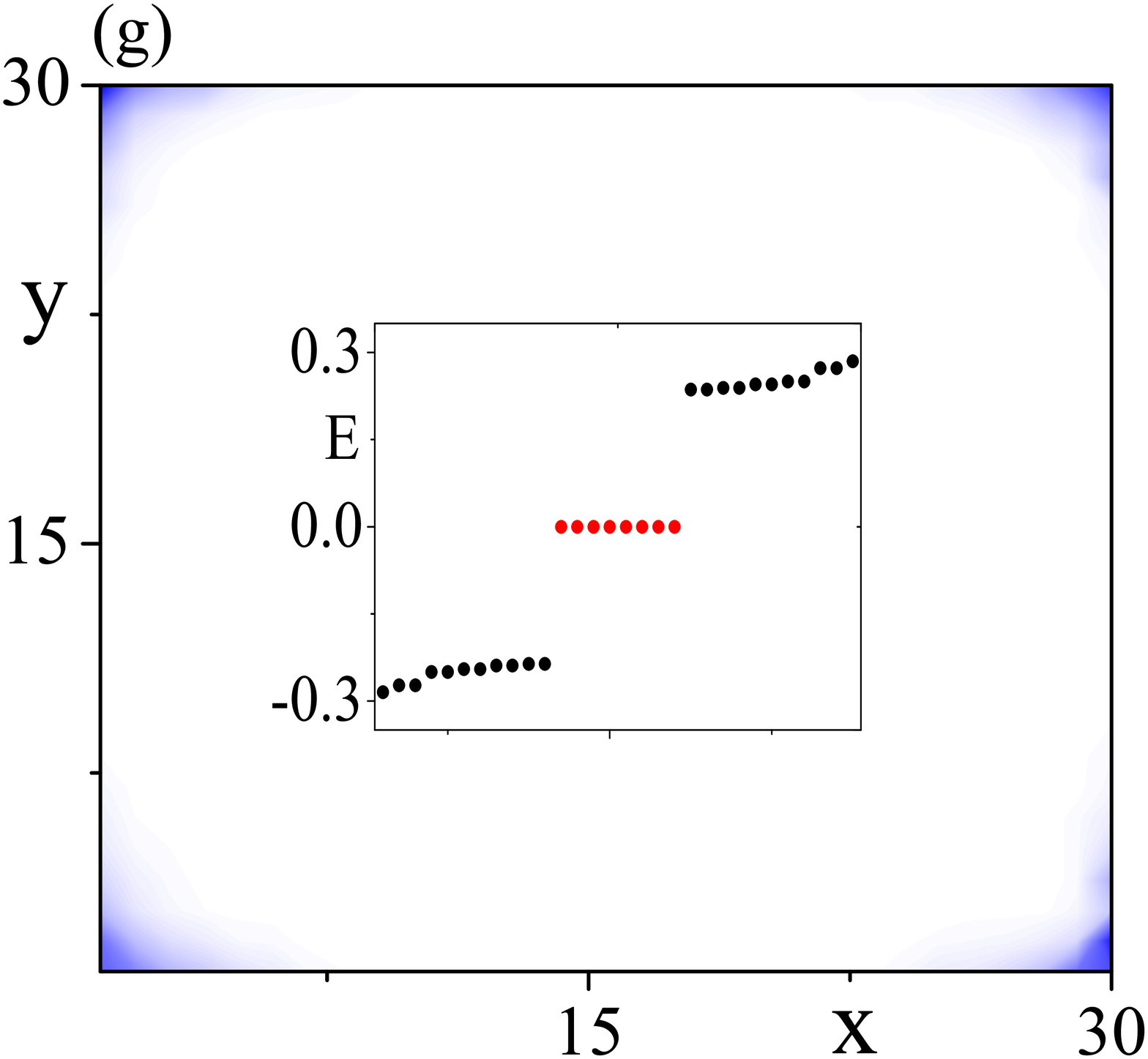}}
\subfigure{\includegraphics[width=4.75cm, height=4cm]{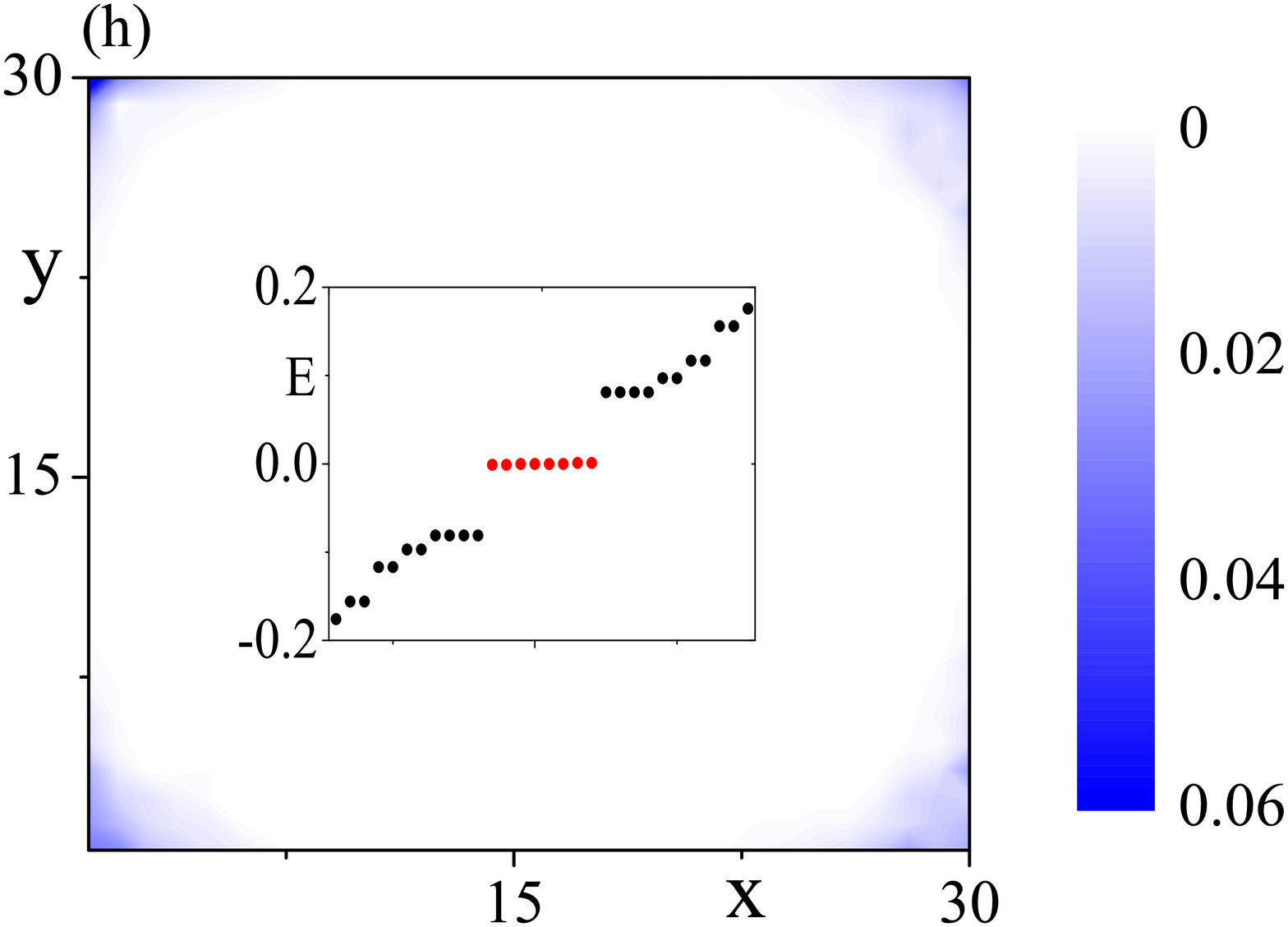}}
\caption{ Upper panels: Profiles of Gaussian random potential with different disorder strengthes (a) $D_{0}=0.2$, (b) $D_{0}=0.4$, (c) $D_{0}=0.6$, (d) $D_{0}=0.8$. Other parameters are
$m_{0}=1.5$, $t_{x}=t_{y}=1.0$, $A_{x}=A_{y}=1.0$, $\Delta_{0}=0$, $\Delta_{x}=-\Delta_{y}=0.5$, and $\mu=0$. Lower panels:
(e)(f)(g)(h) shows the energies and density profiles of the Majorana corner modes of
(a)(b)(c)(d), respectively. The zero-energy corner modes are robust against the disorders. }\label{disorder}
\end{figure*}

To realize a single Majorana zero mode instead of a Majorana Kramers pair per corner, time-reversal symmetry
must be broken. We find that it can be achieved by adding an appropriate in-plane magnetic field in either the $d$-wave or the $s_\pm$-wave case, provided that the system is anisotropic. For concreteness, suppose that the magnetic field is added in the $x$ direction, then it introduces a Zeeman energy $V_x s_{x}\tau_{z}$. The low-energy effective theory becomes
\bea H_{\rm edge}=-iA(l)s_{z}\partial_{l}+M(l)s_{y}\tau_{y} +V_{x}s_{x}\tau_{z}. \eea Now we can split this $4\times 4$ Hamiltonian into two decoupled blocks. In fact, in the two-dimensional $\tau_x=\pm s_z$ subspace, we have $V_x s_x\tau_z = -iV_x s_x \tau_x\tau_y = \mp i V_x s_x s_z \tau_y =\mp V_x s_y\tau_y$, therefore, the two decoupled blocks have Dirac mass $M(l)-V_x$ and $M(l)+V_x$, respectively, and single MZM appears when one of them has a sign changing, while the other does not. This is satisfied, for example, when $|\Delta_d m/t_y| <V_x< |\Delta_d m/t_x|$ for the $d$-wave case.

\section{VI. Stability of the Majorana corner modes against disorders}

In this section, we investigate the stability of the Majorana Kramers pairs against random disorders.
For concreteness, we consider random on-site potentials.
The real-space Hamiltonian reads
\begin{widetext}
\begin{eqnarray}
H&=&\sum_{i,j,\alpha,\alpha',\beta,\beta'}m_{0}c_{i,j;\alpha\beta}^{\dag}(\sigma_{z})_{\alpha\alpha'}(s_{0})_{\beta\beta'}c_{i,j;\alpha'\beta'}
-\{\frac{t_{x}}{2}c_{i,j;\alpha\beta}^{\dag}(\sigma_{z})_{\alpha\alpha'}(s_{0})_{\beta\beta'}c_{i+1,j;\alpha'\beta'}+
\frac{t_{y}}{2}c_{i,j;\alpha\beta}^{\dag}(\sigma_{z})_{\alpha\alpha'}(s_{0})_{\beta\beta'}c_{i,j+1;\alpha'\beta'}+h.c.\}\nonumber\\
&&-\{i\frac{A_{x}}{2}c_{i,j;\alpha\beta}^{\dag}(\sigma_{x})_{\alpha\alpha'}(s_{z})_{\beta\beta'}c_{i+1,j;\alpha'\beta'}+
i\frac{A_{y}}{2}c_{i,j;\alpha\beta}^{\dag}(\sigma_{y})_{\alpha\alpha'}(s_{0})_{\beta\beta'}c_{i,j+1;\alpha'\beta'}+h.c.\}
-\mu c_{i,j;\alpha\beta}^{\dag}(\sigma_{0})_{\alpha\alpha'}(s_{0})_{\beta\beta'}c_{i,j;\alpha'\beta'}\nonumber\\
&&-\{i\frac{\Delta_{0}}{2}c_{i,j;\alpha\beta}^{\dag}(\sigma_{0})_{\alpha\alpha'}(s_{y})_{\beta\beta'}c_{i,j;\alpha'\beta'}^{\dag}
+i\frac{\Delta_{x}}{2}c_{i,j;\alpha\beta}^{\dag}(\sigma_{0})_{\alpha\alpha'}(s_{y})_{\beta\beta'}c_{i+1,j;\alpha'\beta'}^{\dag}
+i\frac{\Delta_{y}}{2}c_{i,j;\alpha\beta}^{\dag}(\sigma_{0})_{\alpha\alpha'}(s_{y})_{\beta\beta'}c_{i,j+1;\alpha'\beta'}^{\dag}+h.c.\}\nonumber\\
&&+V_{ij}c_{i,j;\alpha\beta}^{\dag}(\sigma_{0})_{\alpha\alpha'}(s_{0})_{\beta\beta'}c_{i,j;\alpha'\beta'}
\end{eqnarray}
\end{widetext}
where $s_{x,y,z}$ and $\sigma_{x,y,z}$ are Pauli matrices acting on
spin and orbital degree of freedom, respectively, $s_{0}$ and $\sigma_{0}$ are the
$2\times 2$ unit matrix. The last term represents the random on-site potential, which is taken to obey a Gaussian distribution, namely
$\langle V_{ij}\rangle=0$, $\langle V_{ij}V_{nm}\rangle=D_{0}^{2}\delta_{in}\delta_{jm}$ with
$D_{0}$ characterizing the strength of the randomness.

Fig.\ref{disorder} shows several profiles of Gaussian disorders and the spectra close to zero energy. It is apparent that the Majorana Kramers pairs are robust against disorders for a quite broad range of disorder strength.

\begin{figure*}[thb]
\subfigure{\includegraphics[width=5cm, height=5cm]{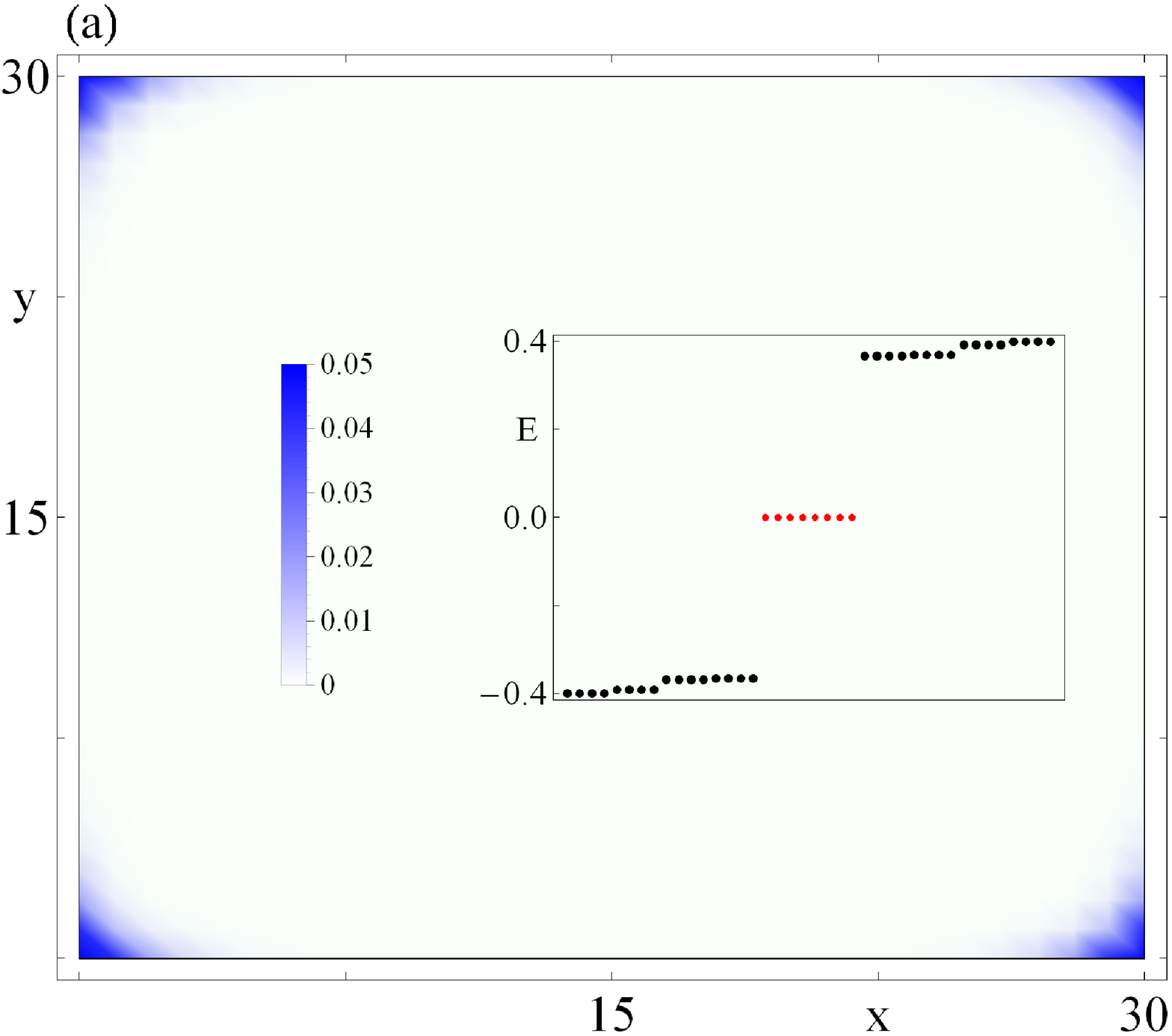}}
\subfigure{\includegraphics[width=5cm, height=5cm]{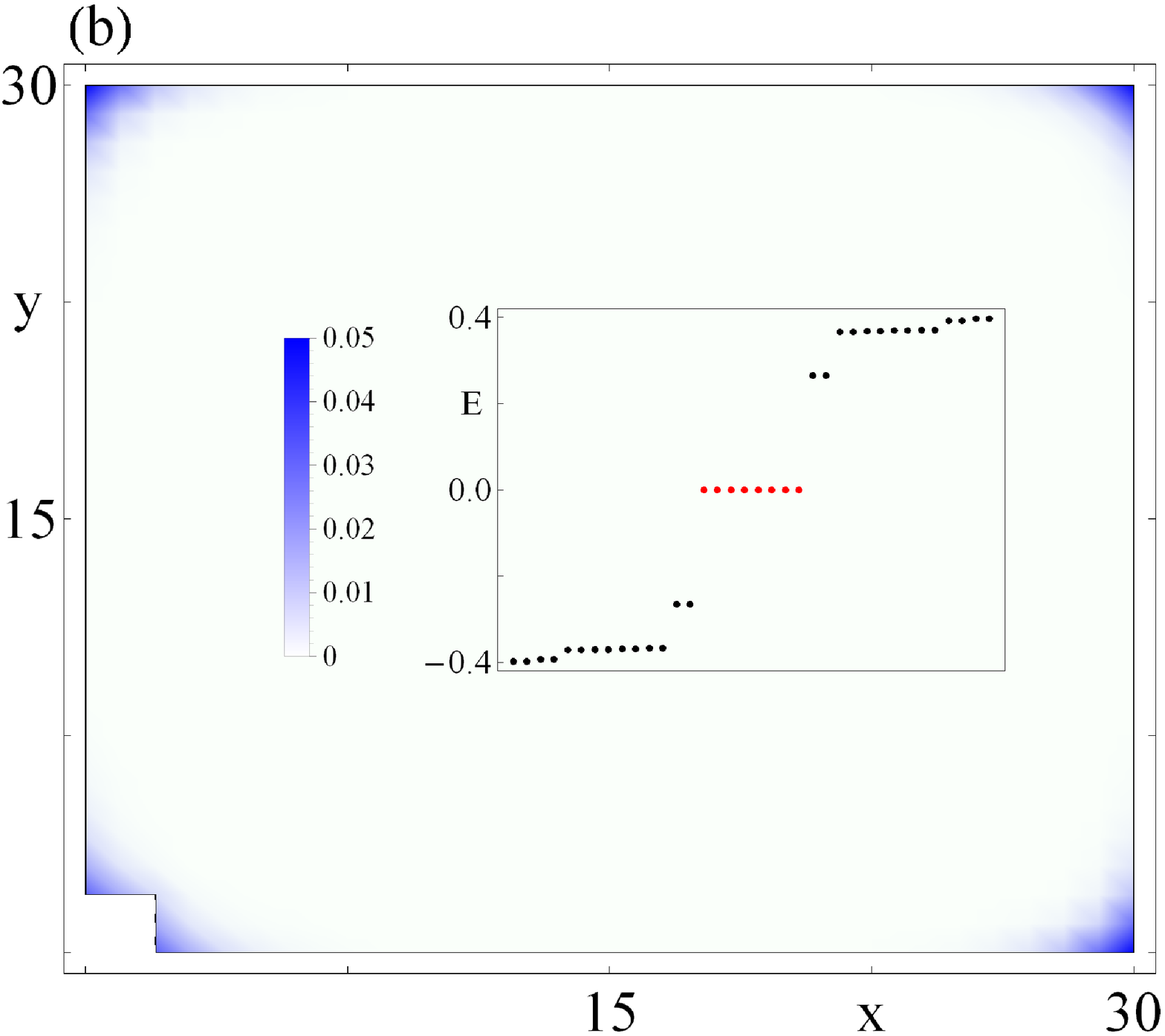}}
\subfigure{\includegraphics[width=5cm, height=5cm]{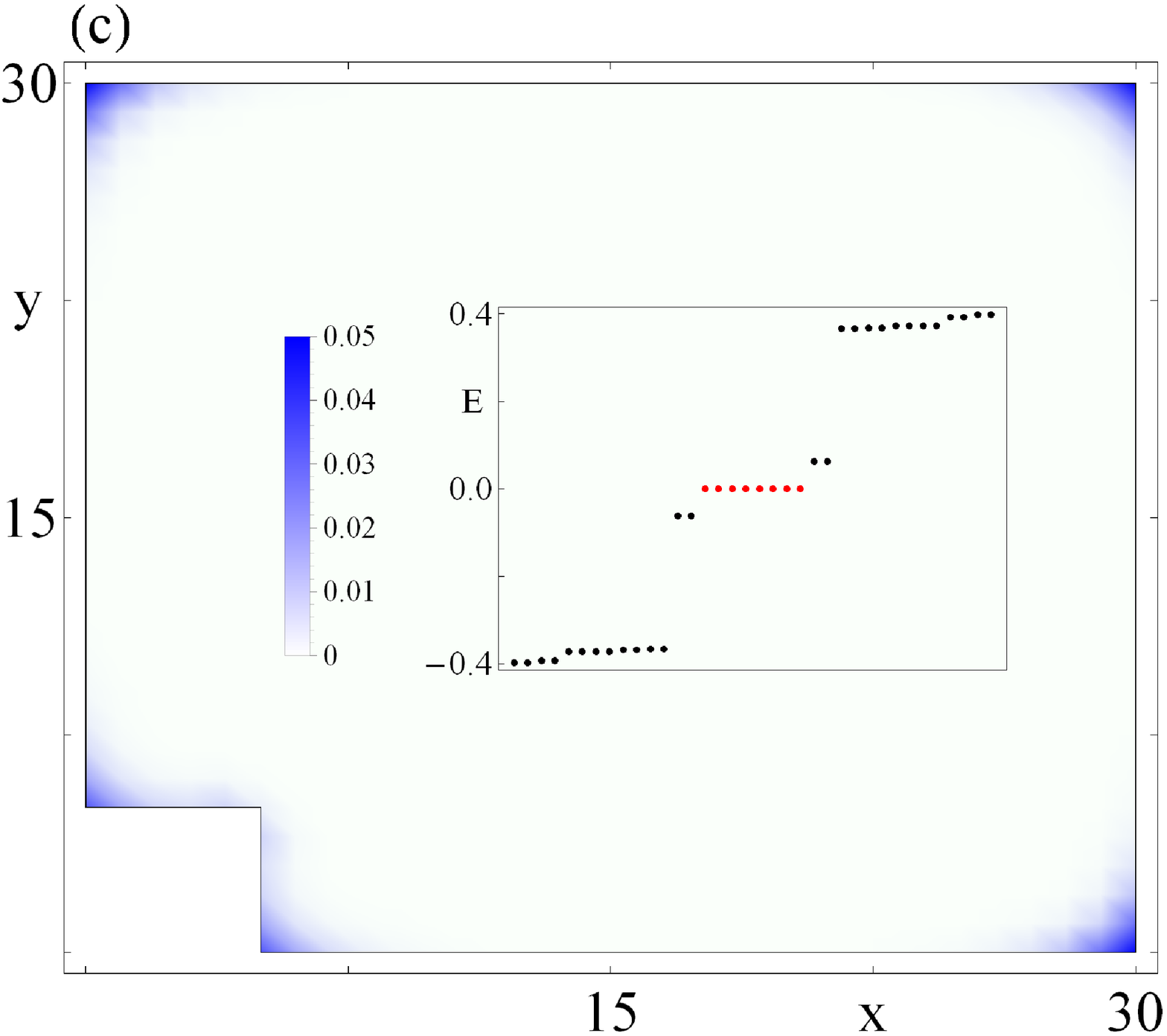}}
\subfigure{\includegraphics[width=5cm, height=5cm]{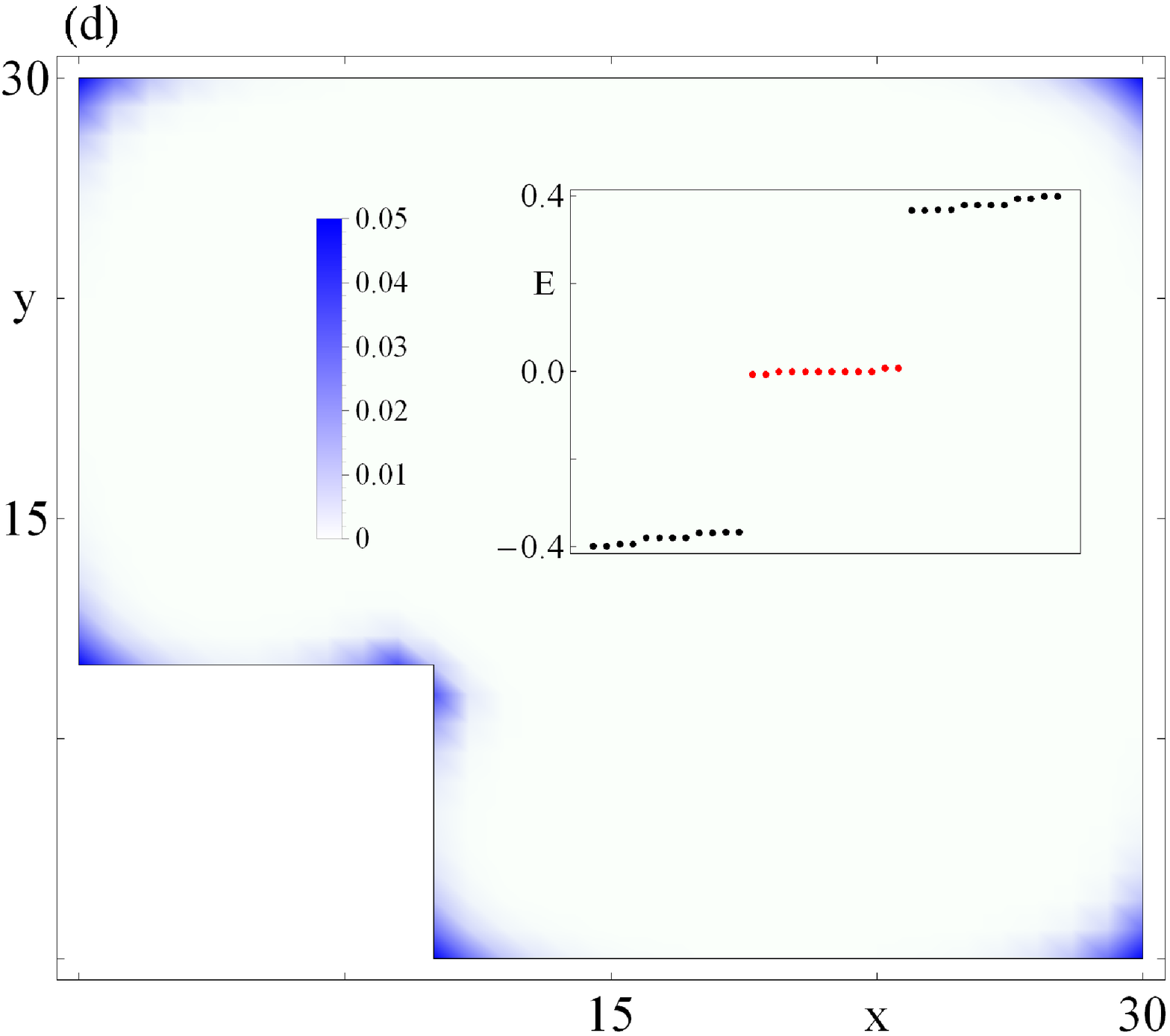}}
\subfigure{\includegraphics[width=5cm, height=5cm]{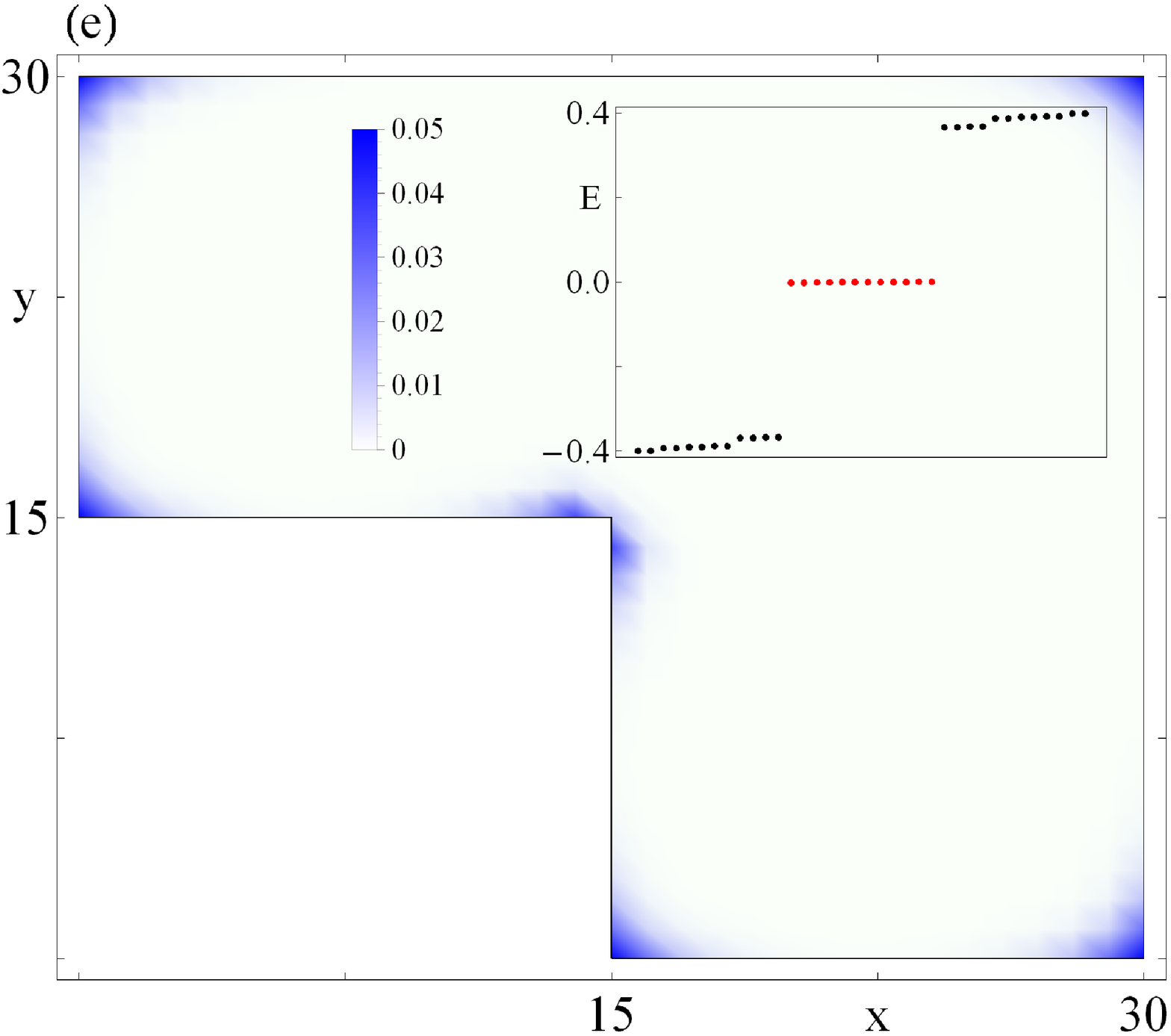}}
\caption{ Majorana Kramers pairs in a non-ideal square lattice with a small square removed at a corner. When the size of removed square is small [panel (b)], the original zero modes remain robust, though the mode profiles are modified. No additional zero modes are created. When the size of removed square is large, it creates additional zero modes [panels (d)(e)].
$t_{x}=t_{y}=1$, $m_{0}=1.2$,
$A_{x}=A_{y}=1$, $\Delta_{0}=0$, $\Delta_{x}=-\Delta_{y}=0.5$, $\mu=0$. Shown in each sample is the local density states at $E=0$ (The Dirac delta function is replaced by a Lorentzian function with a small width $0.03$).  }\label{defect}
\end{figure*}

\section{VII. Effects of edge imperfections }

In real experiments, the edges are usually not atomically precise, and the edge orientations can vary in space. It is useful to study the stability of Majorana Kramers pairs in the presence of these imperfections. Thanks to the protection of particle-hole symmetry, a Majorana Kramers pair of zero modes remains robust in the presence of small or modest imperfections, because shifting the energy of a spatially isolated Majorana Kramers pair away from zero is inconsistent with the inherent particle-hole symmetry of BdG equation.

As a representative example of edge imperfections, we calculate the energies and eigenstates on a non-ideal square lattice with a small square removed at a corner (see Fig.\ref{defect}). In Fig.\ref{defect}(b), additional in-gap modes are created with energies far away from zero. Such nonzero-energy modes do not affect detection of the original Majorana Kramers pairs, which remain at zero energy. When the size of the removed square is increased [Fig.\ref{defect}(c)], the energies of in-gap modes are lowered, and finally, their energies come to zero, namely, additional Majorana Kramers pairs are created [Fig.\ref{defect}(d)(e)]. From this example, we can see that lattice imperfections generally do not destroy the Majorana Kramers pairs; instead, they provide more opportunity to detect them: The imperfections provide additional corners built in the samples, hosting additional Majorana Kramers pairs.

\section{VIII. Effects of chemical potential }

\begin{figure*}[htb]
\subfigure{\includegraphics[width=4cm, height=3.5cm]{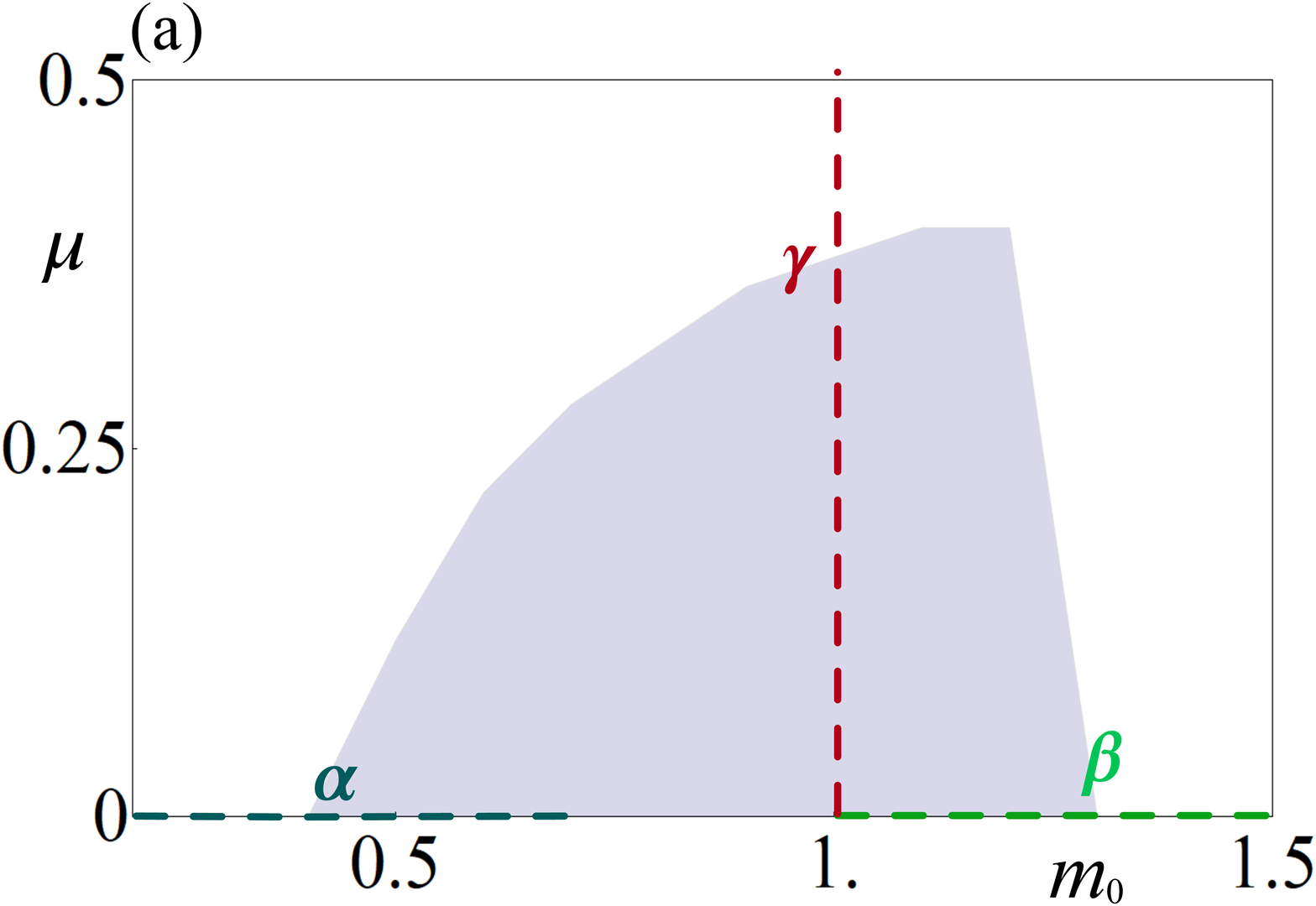}}
\subfigure{\includegraphics[width=4cm, height=3.5cm]{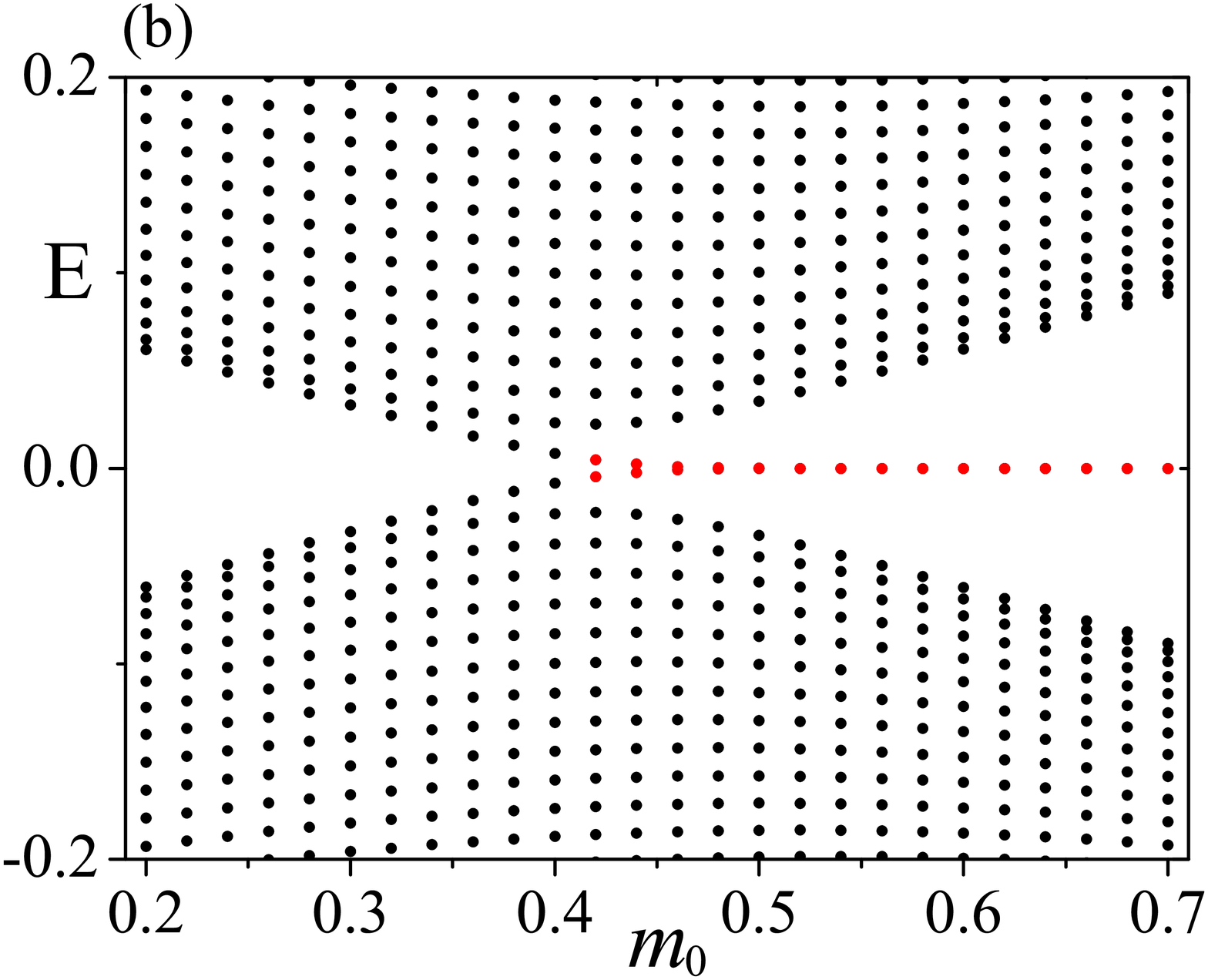}}
\subfigure{\includegraphics[width=4cm, height=3.5cm]{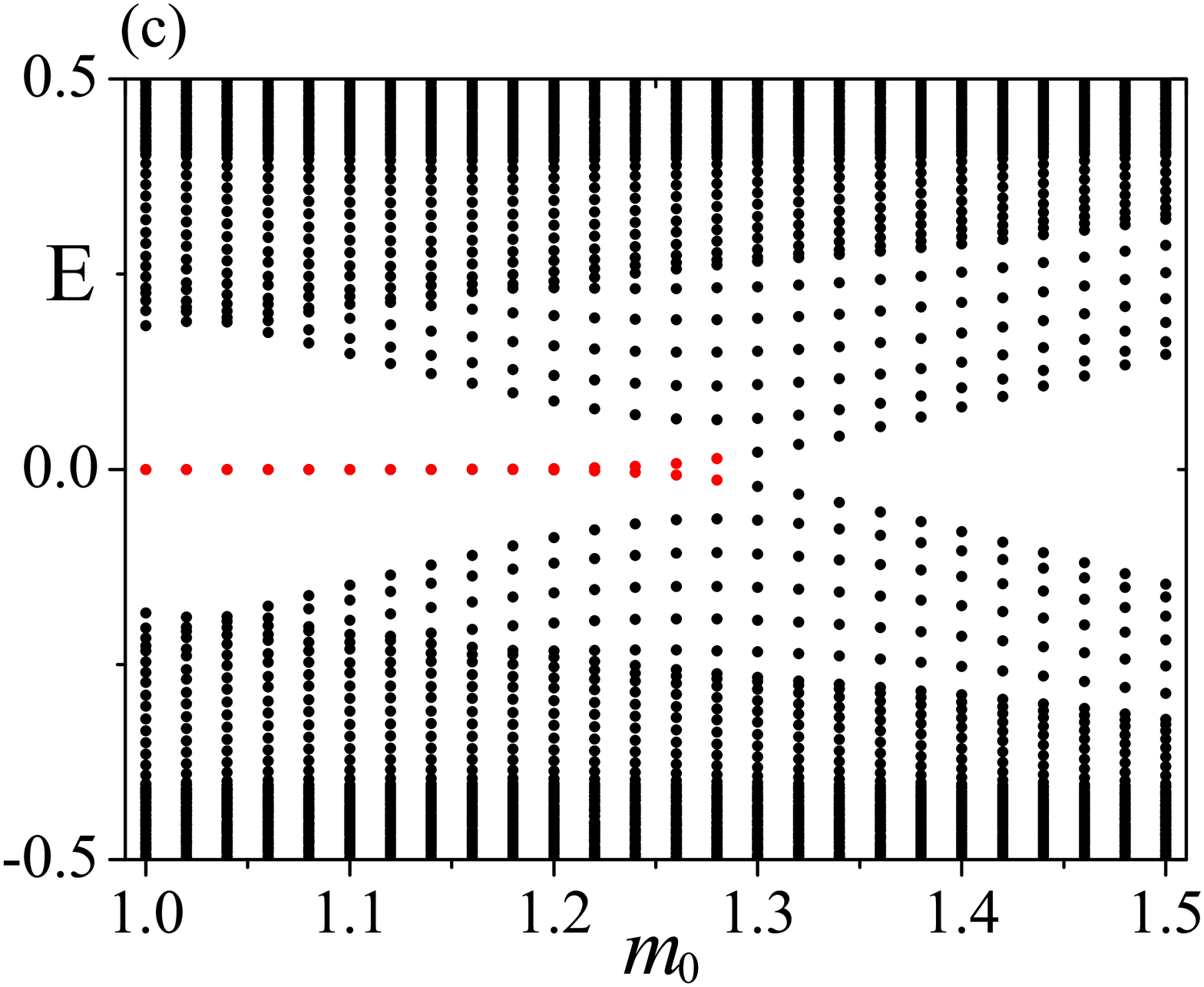}}
\subfigure{\includegraphics[width=4cm, height=3.5cm]{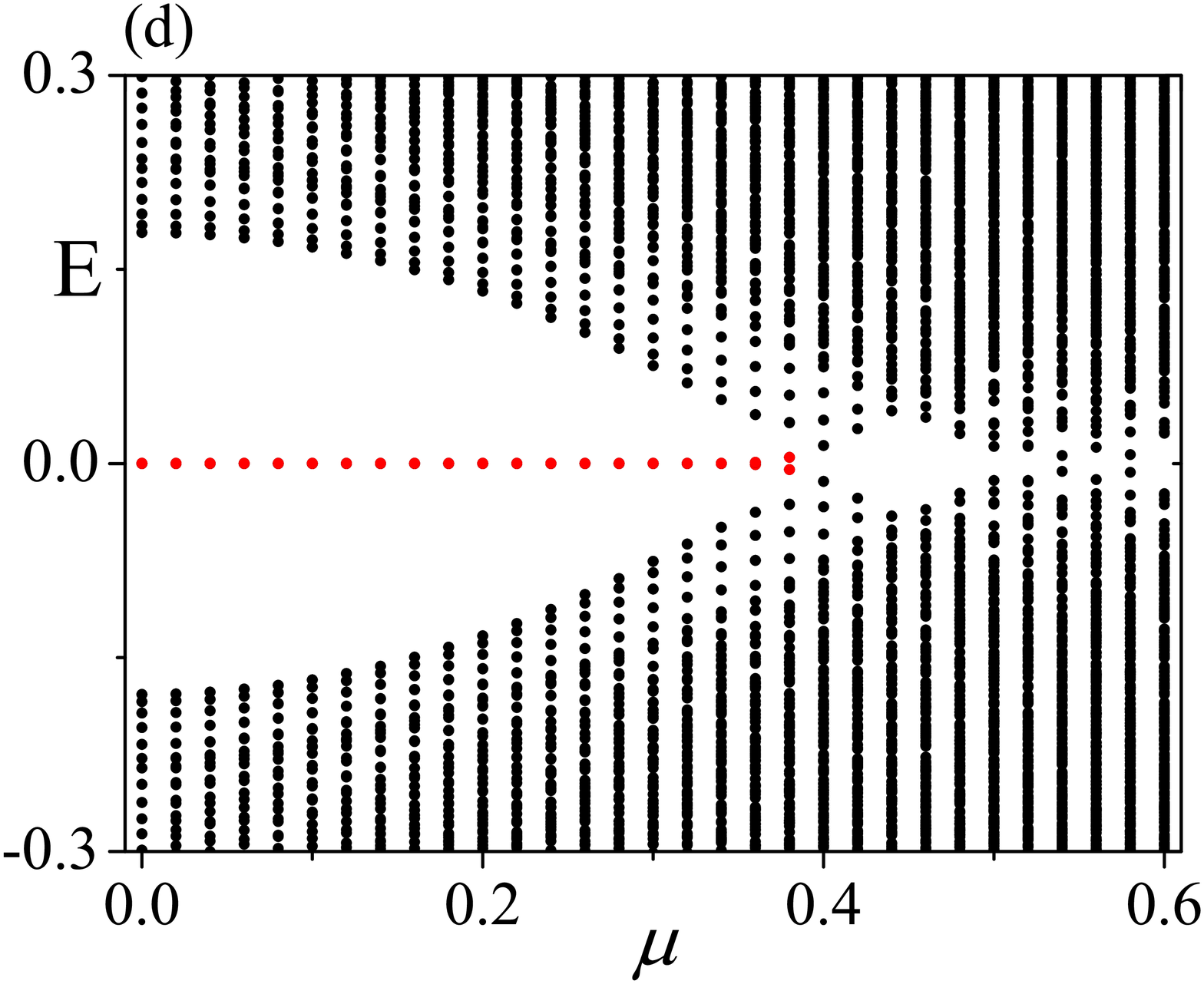}}
\caption{ (a) A cross section of the 3D phase diagram in Fig.4(d) of the main text. The cross section corresponds to $\Delta_{0}=\Delta_{1}$.
In the shadow region the Majorana corner modes are found to exist.  (b)(c)(d) show the energy spectrum when the parameters change along
the path $\alpha$, $\beta$, $\gamma$ shown in (a), respectively.
The system size is (b) $L_{x}\times L_{y}=80\times20$,
(c) $L_{x}\times L_{y}=20\times90$, (d) $L_{x}\times L_{y}=40\times40$.
Common parameters are $t_{x}=A_{x}=0.4$, $t_{y}=A_{y}=1.3$, $\Delta_{0}=\Delta_{1}=0.4$.
   }\label{tdpd}
\end{figure*}

In this section, we provide more details about the effects of a nonzero chemical potential. We will see that including the chemical potential does not change the physics qualitatively compared to the case of vanishing chemical potential.

First, we study the effect of chemical in the low-energy effective edge theory.  At the edge I, nonzero $\mu$ simply generates the following term in the effective edge Hamiltonian:
\begin{eqnarray}
(\Delta H)_{\alpha\beta}=-\int_{0}^{+\infty}dx\psi^*_{\alpha}(x)\mu\tau_{z}\psi_{\beta}(x)=-\mu(\tau_{z})_{\alpha\beta}.
\end{eqnarray}
Now the edge effective Hamiltonian reads
\begin{eqnarray}
H_{\rm edge}=-iA(l)s_{z}\partial_{l}-\mu\tau_{z}+M(l)s_{y}\tau_{y}.
\end{eqnarray}
This Hamiltonian can be decomposed as the sum of two independent $2\times 2$ Hamiltonians
\begin{eqnarray}
H_{\rm edge,-}&=&-iA(l)s_{z}\partial_{l}-\mu s_{z}-M(l)s_{x},\nonumber\\
H_{\rm edge,+}&=&-iA(l)s_{z}\partial_{l}+\mu s_{z}+M(l)s_{x},
\end{eqnarray} where $H_{\rm edge,-}$ acts in the $\tau_z=s_z$ subspace spanned by $\{|s_{z}=1,\tau_{z}=1\rangle, |s_{z}=-1,\tau_{z}=-1\rangle\}$, and $H_{\rm edge,+}$ acts in the $\tau_z=-s_z$ subspace spanned by $\{|s_{z}=1,\tau_{z}=-1\rangle, |s_{z}=-1,\tau_{z}=1\rangle\}$. Now the zero modes can be found as
\begin{eqnarray}
|\psi^{\pm }(l)\rangle\propto e^{-\int^l dl' [M(l')\mp i\mu]/A(l')}.
\end{eqnarray} Here, $\mu$ only adds trivial phase factors to the wavefunctions without modifying their profiles. It indicates that a small or modest $\mu$ (to which the low-energy theory is applicable) does not qualitatively change the physics.

Now we go beyond the low-energy continuum Hamiltonian to the lattice Hamiltonian. As we increase the chemical potential $\mu$ away from $0$, there are two possible scenarios that can kill the Majorana Kramers pairs: (i) The bulk gap closes; (ii) the edge gap closes. First, let us consider the first scenario. The BdG Hamiltonian of the bulk is
\begin{eqnarray}
H(\bk)=&&M(\bk)\sigma_{z}\tau_{z}+A_{x}\sin k_{x}\sigma_{x}s_{z}+A_{y}\sin k_{y}\sigma_{y}\tau_{z} \nn \\ && +\Delta(\bk)s_{y}\tau_{y}-\mu\tau_z,
\end{eqnarray}
whose energy spectra are
\begin{eqnarray}
E(\bk)=\pm\sqrt{(\sqrt{M^{2}(\bk)+A_{x}^{2}\sin^{2}k_{x}+A_{y}^{2}\sin^{2}k_{y}}\pm\mu)^{2}+\Delta^{2}(\bk)}.\nonumber
\end{eqnarray}
Without the pairing, the gap closing condition is $\sqrt{M^{2}(\bk)+A_{x}^{2}\sin^{2}k_{x}+A_{y}^{2}\sin^{2}k_{y}}=|\mu|$, which determines the bulk Fermi surface of the doped TI. Without losing generality, we only focus on the $\mu>0$ case. The Fermi surface appears when $\mu>\text{min}\{\sqrt{M^{2}(\bk)+A_{x}^{2}\sin^{2}k_{x}+A_{y}^{2}\sin^{2}k_{y}}\}$.  When the Fermi surface determined by \bea \sqrt{M^{2}(\bk)+A_{x}^{2}\sin^{2}k_{x}+A_{y}^{2}\sin^{2}k_{y}}=\mu \eea
and the pairing nodal ring determined by $\Delta(\bk)=0$ cross each other, the energy gap closes.
For the pairing we considered in the main text, $\Delta(\bk)=\Delta_{0}-\Delta_{1}(\cos k_{x}+\cos k_{y})$ with
$0<\Delta_{0}<2\Delta_{1}$, we find the gapless region  is \bea \mu_{1}<\mu<\mu_{2}, \eea where
$\mu_{1}=\text{min}\{\sqrt{(m_{0}-t_{x}\frac{(\Delta_{0}-\Delta_{1})} {\Delta_{1}}-t_{y})^{2}+A_{x}^{2}(1-\frac{(\Delta_{0}-\Delta_{1})^{2}}{\Delta_{1}^{2}})}$,
$\sqrt{(m_{0}-t_{x}-t_{y}\frac{(\Delta_{0}-\Delta_{1})}{\Delta_{1}})^{2}+A_{y}^{2}(1-\frac{(\Delta_{0}-\Delta_{1})^{2}}{\Delta_{1}^{2}})}\}$, and
$\mu_{2}=\text{max}\{\sqrt{(m_{0}-t_{x}\frac{(\Delta_{0}-\Delta_{1})} {\Delta_{1}}-t_{y})^{2}+A_{x}^{2}(1-\frac{(\Delta_{0}-\Delta_{1})^{2}}{\Delta_{1}^{2}})},$
$\sqrt{(m_{0}-t_{x}-t_{y} \frac{(\Delta_{0}-\Delta_{1})}{\Delta_{1}})^{2}+A_{y}^{2}(1-\frac{(\Delta_{0}-\Delta_{1})^{2}}{\Delta_{1}^{2}})}\}$.

The scenario (ii) mentioned above, namely closing the edge gap to kill the Majorana Kramers pairs, is not as convenient to quantify. The reason is that the edge states of TI do not exist as a 1D system by itself, and the edge cannot be separately studied beyond the low-energy theory. Nevertheless, we can numerically find the critical chemical potential $\mu_c$ above which the Majorana Kramers pairs disappear. When the Cooper pairing is small, we find that $\mu_{c}<\mu_{1}$ in general, which indicates that the gap closing appear at the edge, i.e., the scenario (ii).

A 3D phase diagram with varied $(m_0, \Delta_0/\Delta_1, \mu)$ is numerically calculated and shown in Fig.4(d) in the main article.
In this supplemental material, let us focus on the the $\Delta_{0}/\Delta_{1}=1$ cross section of this 3D phase diagram, which is shown in Fig.\ref{tdpd}(a). For the chosen parameters, the paring nodal ring and the band-inversion ring cross each other when $0.4<m_{0}<1.3$. From Fig.\ref{tdpd}(b)(c), it is clear that zero modes are indeed found when $0.4\lesssim m_0 \lesssim 1.3$,
in accordance with the criterion Eq.(19) of the main article.
Along the path $\gamma$, Majorana corner mdoes exist for $\mu \lesssim 0.38$ (Fig.\ref{tdpd}(d)). Note that the bulk gap closes at $\mu_{1}=\sqrt{(m_{0}-1.3)^{2}+0.4^{2}}=0.5$, therefore, the disappearance of Majorana corner modes is due to the gap closing at the edges.

\end{document}